\documentclass[aps,prd,twocolumn,showpacs,floatfix,preprintnumbers,amsmath,amssymb,nofootinbib,groupedaddress]{revtex4-1}

\input epsf
\usepackage{graphicx}
\usepackage{color}
\usepackage{mathtools}

\def\VEV#1{\left\langle #1 \right\rangle}

 \newcommand{\f}{f_{\mathrm{NL}}}

    \newcommand{\be}{\begin{equation}}
  \newcommand{\ee}{\end{equation}}

\newcommand{\fsky}{f_{\rm sky}}
\newcommand{\tto}{\overline{T}_{21}}

\newcommand{\bsx}{\boldsymbol{x}}
\newcommand{\bsk}{\boldsymbol{k}}
\newcommand{\bsl}{\boldsymbol{\ell}}
\newcommand{\bs}{\boldsymbol}
\newcommand{\rme}{\textrm{e}}
\newcommand{\beq}{\begin{equation}}
\newcommand{\eeq}{\end{equation}}
\newcommand{\barr}{\begin{eqnarray}}
\newcommand{\earr}{\end{eqnarray}}

\begin{document}

\title{Primordial non-gaussianity from the bispectrum of 21-cm fluctuations in the dark ages}

\author{Julian B. Mu\~noz} \email{julianmunoz@jhu.edu}
\author{Yacine Ali-Ha\"imoud} \email{yacine@jhu.edu}
\author{Marc Kamionkowski} \email{kamion@jhu.edu}

\affiliation{Department of Physics and Astronomy, Johns
     Hopkins University, 3400 N.\ Charles St., Baltimore, MD 21218}

\date{\today}

\begin{abstract}
A measurement of primordial non-gaussianity will be of paramount importance to distinguish between different models of inflation. Cosmic microwave background (CMB) anisotropy observations have set unprecedented bounds on the non-gaussianity parameter $\f$ but the interesting regime $\f \lesssim 1$ is beyond their reach. Brightness-temperature fluctuations in the 21-cm line during the dark ages ($z \sim 30 - 100$) are a promising successor to CMB studies, giving access to a much larger number of modes. They are, however, intrinsically non-linear, which results in secondary non-gaussianities orders of magnitude larger than the sought-after primordial signal. In this paper we carefully compute the primary and secondary bispectra of 21-cm fluctuations on small scales. We use the flat-sky formalism, which greatly simplifies the analysis, while still being very accurate on small angular scales. We show that the secondary bispectrum is highly degenerate with the primordial one, and argue that even percent-level uncertainties in the amplitude of the former lead to a bias of order $\Delta f_{\rm NL} \sim 10$. To tackle this problem we carry out a detailed Fisher analysis, marginalizing over the amplitudes of a few smooth redshift-dependent coefficients characterizing the secondary bispectrum. We find that the signal-to-noise ratio for a single redshift slice is reduced by a factor of $\sim 5$ in comparison to a case without secondary non-gaussianities.  
Setting aside foreground contamination, we forecast that a cosmic-variance-limited experiment observing 21-cm fluctuations over $30 \leq z \leq 100$ with a 0.1-MHz bandwidth and 0.1-arcminute angular resolution could achieve a sensitivity of order $\f^{\mathrm{local}} \sim 0.03$, $\f^{\mathrm{equil}} \sim 0.04$ and $\f^{\mathrm{ortho}} \sim 0.03$.

\end{abstract}


\maketitle

\section{Introduction}
\label{sec:intro}

Increasingly precise cosmic microwave background (CMB) \cite{1212.5226, 1502.02114} and large-scale structure \cite{1501.03851,0902.4759} measurements have zeroed in on a rather simple model of the cosmos, requiring only a handful of parameters. In particular, initial fluctuations seem to be mostly scalar and highly gaussian \cite{1403.3985,1502.00612,1403.2369}. They are well described by a simple power-law spectrum, whose slope is consistent with a single scalar field driving inflation while slowly rolling down a very flat potential \cite{0907.5424,Linde:1983gd}. Proposed experiments like EUCLID \cite{1206.1225} and PRISM \cite{1310.1554} will measure the inflationary parameters even more precisely, and possibly extract additional quantities, such as the running of the scalar tilt \cite{1007.3748}. This should further constrain the form of the inflaton potential during the quasi-de-Sitter phase.

While single-field inflation has the merit of simplicity, a plethora of alternative models remain consistent with current data \cite{0902.4731,1007.0027,1106.1428,astro-ph/9804177,astro-ph/9904309}. The main characteristic that differentiates them from the simplest inflationary scenario is that they can generate significant primordial non-gaussianities (PNGs). The simplest form of PNG is a non-vanishing three-point function for the primordial curvature perturbation $\zeta$, parametrized by a dimensionless amplitude $\f \sim \langle \zeta^3 \rangle/\langle \zeta^2\rangle^2$. Single-field inflation leads to a small three-point function, corresponding to $\f \sim 10^{-2}$ \cite{astro-ph/0210603,astro-ph/0407059}. Alternative models typically generate $\f \sim 1$, as a result of interactions with other fields \cite{Linde_97,hep-th/0507205}, higher-derivative terms in the Lagrangian \cite{hep-th/0404084,hep-th/0605045,hep-th/9904075}, or other mechanisms \cite{0902.4731}. Measuring $\f \lesssim 1$ is therefore a natural target for future experiments to start significantly constraining the physics of inflation \cite{1412.4671}.

The best constraints on $\f$ to date are obtained from CMB studies \cite{1502.01592}, and are consistent with zero, though with a large uncertainty, $\sigma_{\f} \sim 5 - 40$ depending on the shape considered. CMB measurements are now cosmic-variance limited in temperature down to the photon diffusion scale corresponding to multipole $\ell \sim 2000$. The anticipated improvement in polarization measurements is expected to only marginally tighten the constraints on $\f$. Reaching the $\f \sim 1$ frontier will therefore most likely require other data sets.

Fluctuations in the brightness temperature of the 21-cm line of neutral hydrogen have the potential to open a new window on the high-redshift universe \cite{astro-ph/0610257,astro-ph/0611126}. This observable can in principle allow us to probe a fantastic number of modes, largely surpassing those available from CMB observations alone. First, fluctuations are undamped down to the baryon Jeans scale (with wavenumber $k \sim 300$ Mpc$^{-1}$) three orders of magnitude smaller than the photon diffusion scale ($k \sim 0.2$ Mpc$^{-1}$). In addition, whereas CMB anisotropies probe a single surface, a line such as the 21-cm transition makes it possible to observe the early universe in tomography, and to co-add the information from each independent redshift slice. While the 21-cm line can in principle be observed all the way to cosmological reionization, at $z \sim 10$, the signal is cleaner at higher redshifts $z \gtrsim 30$, the dark ages preceding the formation of the first luminous objects. We focus on this redshift range in this paper.

The technical challenges to observe high-redshift 21-cm fluctuations are daunting, and will most likely require a telescope array on the far side of the Moon \cite{0902.0493}, as well as foreground-removal of the Galactic synchrotron radiation to an exquisite accuracy. We will not tackle these problems in the present work, but focus on another source of contamination: the intrinsic non-gaussian nature of 21-cm fluctuations, even for perfectly gaussian initial conditions.

While the baryon-photon fluctuations are highly linear at the epoch of last-scattering $z \sim 1100$, the perturbations in the cold dark matter (CDM) and baryon fluids have significantly grown by $z \sim 50$. Even if they remain small enough that no bound structure has formed yet, gravitational growth leads to a non-linear dependence of the density field on initial conditions. In addition, the 21-cm brightness temperature depends non-linearly on the local baryon density, velocity gradient, and temperature. Both effects give rise to a non-vanishing three-point function for the 21-cm brightness temperature, orders of magnitude larger than that resulting from PNG. Unless treated appropriately, this can jeopardize the usefulness of 21-cm fluctuations to measure PNG.

Two previous studies have partially addressed this issue. Ref.~\cite{astro-ph/0610257} computed the bispectrum of 21-cm fluctuations resulting from non-linear gravitational growth, but treated it approximately as a confusion noise rather than a bias. Ref.~\cite{astro-ph/0611126} computed all contributions of the secondary bispectrum, but did not account for it in their final forecasts. In addition they only computed the bispectrum for specific triangle configurations. These two groups moreover get significantly different final results.  
 
In this paper we compute the primary and secondary bispectra using the flat-sky formalism. This accurately reproduces the full-sky calculation with a much lower computational cost, and greatly simplifies the analysis. We show that the shapes of the primary and secondary bispectra overlap significantly. Unsubtracted, the secondary bispectrum would lead to a bias $\Delta \f \sim 10^3$. Even percent-level residuals after subtraction would lead to a non-zero non gaussianity of order $\Delta \f \sim 10$. This warrants a Fisher analysis, fitting simultaneously for the amplitude of PNG and for nuisance parameters characterizing the residual secondary bispectrum after a best-estimate is subtracted. For a single redshift slice, we find that the uncertainty in $\f$ after marginalizing over the nuisance parameters is increased by a factor of $\sim 3-6$ in comparison to an ideal case without secondaries. Finally, we optimally combine redshift slices accounting for the smoothness of the secondary bispectrum as a function of redshift. Our forecasts for a cosmic-variance-limited experiment targeting $30 \leq z \leq 100$ with a bandwidth of 0.1 MHz and angular resolution of 0.1 arcminute are: 
 $\sigma_{\f^{\mathrm{local}}} \sim 0.03$, $\sigma_{\f^{\mathrm{equil}}} \sim 0.04$, and $\sigma_{\f^{\mathrm{ortho}}} \sim 0.03$. For the same angular resolution but a bandwidth of 1 MHz our forecast is  $\sigma_{\f^{\mathrm{local}}} \sim 0.12$, $\sigma_{\f^{\mathrm{equil}}} \sim 0.39$, and $\sigma_{\f^{\mathrm{ortho}}} \sim 0.29$.
 
The paper is structured as follows, in Section II, we briefly review the basic physics of the 21-cm transition and the flat-sky formalism. In Section III, we compute the different sources of non-gaussianities, both primordial and secondary. In Section IV, we forecast the potential signal-to-noise ratio reachable for cosmic-variance limited experiments. We conclude in Section V.

\section{Basic assumptions and notation}

\subsection{21-cm brightness temperature}

The \emph{spin temperature} of neutral hydrogen $T_s$ is defined as usual through the ratio of the abundance of atoms in the triplet state and in the singlet state:
\begin{equation}
\dfrac{n_1}{n_0} = 3~ e^{-T_*/T_s}, 
\end{equation}
where $T_* = 0.068$ K $= 5.9~ \mu$eV is the energy difference between the two hyperfine levels. The abundances $n_0$ and $n_1$ can be obtained to high accuracy in the steady-state approximation by equating the rate of upward and downward transitions: 
\beq
n_0 (C_{01} + R_{01}) = n_1 (C_{10} + R_{10}),
\eeq 
where the $C_{ij}$ are the collisional transition rates, proportional to the gas density and dependent on the gas temperature $T_{\rm gas}$, and the $R_{ij}$ are the rates of radiative transitions mediated by CMB blackbody photons (specifically, $R_{10}$ includes spontaneous and stimulated emission, and $R_{01}$ accounts for absorption). Since we are concerned with the dark ages preceding the formation of the first luminous objects, we do not account for transitions resulting from inelastic scattering of Lyman-$\alpha$ photons (the Wouthuysen-Field effect \cite{Wouthuysen,Field,astro-ph/0507102}). 
The steady-state approximation is very accurate as $C_{ij} + R_{ij} \gg H$ at all times \cite{astro-ph/0702600}. In the limit $T_* \ll T_{\rm gas}, T_{\rm cmb}$, valid at all times, the spin temperature is then given by
\beq
T_s = T_{\rm cmb} + \frac{C_{10}}{C_{10} + A_{10} \frac{T_{\rm gas}}{T_*}} (T_{\rm gas} - T_{\rm cmb}),
\eeq
where $A_{10}$ is the Einstein-A coefficient of the hyperfine transition.

The brightness temperature (or more accurately, the brightness temperature contrast with respect to the CMB) resulting from the resonant interaction of CMB photons with the hyperfine transition is given by
\begin{equation}
T_{21}^{\rm loc} = (T_{s} - T_{\rm cmb}) (1-e^{-\tau}),
\label{eq:Tb0}
\end{equation}
where the superscript ``loc" emphasizes that this is the local brightness temperature, at the location of the absorbing gas.

The optical depth $\tau$ is a function of the local neutral hydrogen density $n_{\rm H^0} = n_{\rm H}(1-x_e)$ (where $x_e$ is the free electron fraction) and the gradient $\partial_r v_r$ of the peculiar velocity along the direction of propagation \cite{Bharadwaj_2004}:
\begin{equation}
\tau = \frac{3}{32 \pi} \frac{T_*}{T_s} n_{\rm H^0} \lambda_{21}^3 \frac{A_{10}}{H(z) + (1+z) \partial_r v_r},
\label{eq:tau}
\end{equation}
where $\lambda_{21} \approx 21$ cm is the transition wavelength. The second term in the denominator accounts for the perturbation of the optical depth due to the perturbed expansion rate along the direction of propagation\footnote{This term is often mislabeled as a ``redshift-space distortion". It indeed has the same form in the optically thin limit, but has a qualitatively different origin \cite{1109.4612}. On the one hand, redshift-space distortions arise from the Jacobian of the transform from real to redshift space to which the observer has access. On the other hand the local velocity gradient modifies the local expansion rate and hence the optical depth (or escape probability) of 21-cm photons, regardless of the relative velocity between the observer and the emitter. This effect is analogous to the perturbation of the Lyman-$\alpha$ escape probability which is one of the sources of perturbed recombination \cite{astro-ph/0702600,0812.3652}.}. The 21-cm transition is optically thin ($\tau \ll 1$) in the regime of interest, so that $T_{21}^{\rm loc} \approx \tau (T_s - T_{\rm cmb})$. The brightness temperature observed today is then just
\beq
T_{21}^{\rm obs} = \frac{T_{21}^{\rm loc}}{1+z} = \tau \frac{T_s - T_{\rm cmb}}{1+z} .
\eeq

\subsection{Fluctuations of the 21-cm brightness temperature}

We define $\delta_v \equiv -(1+z) (\partial_r v_r)/H(z)$. Up to terms of third order in the fluctuations, the observed brightness temperature (we shall drop the superscript ``obs") takes the form \cite{1312.4948}\footnote{Our definition of $\delta_v$ differs from that of Ref.~\cite{1312.4948} by a minus sign.}
\barr
T_{21} = \overline{T}_{21} (1 + \delta_v + \delta_v^2)
+ \left(\mathcal{T}_{b}~ \delta_{b} + \mathcal{T}_{T}
~  \delta_{T_{\rm gas}}\right)(1 + \delta_v) \nonumber\\
+  \mathcal{T}_{bb} ~\delta_{b}^2 +\mathcal{T}_{bT}~
\delta_{b} \delta_{T_{\rm gas}} + \mathcal{T}_{TT} ~\delta_{T_{\rm gas}}^2 ,~\label{eq:T21-full}
  \earr
where $\delta_b \equiv \delta n_b/\overline{n}_b$ is the fractional fluctuation of the baryon density and $\delta_{T_{\rm gas}}$ is the fractional fluctuation of the gas temperature, which affect $T_{21}$ through the collision rates. This equation neglects fluctuations of the ionization fraction $x_e \sim 10^{-4}$ at the redshifts of interest, as they lead to negligible fluctuations of $T_{21}$ which is proportional to $(1 - x_e)$. We compute the coefficients in the above equation as described in Ref.~\cite{1312.4948}. They ought to be used for detailed prediction when actual data is available. For this study, however, we shall make simplifying assumptions regarding the gas temperature fluctuations in order to keep calculations tractable. We now describe our approximations.

\begin{figure}
\includegraphics[width=88mm]{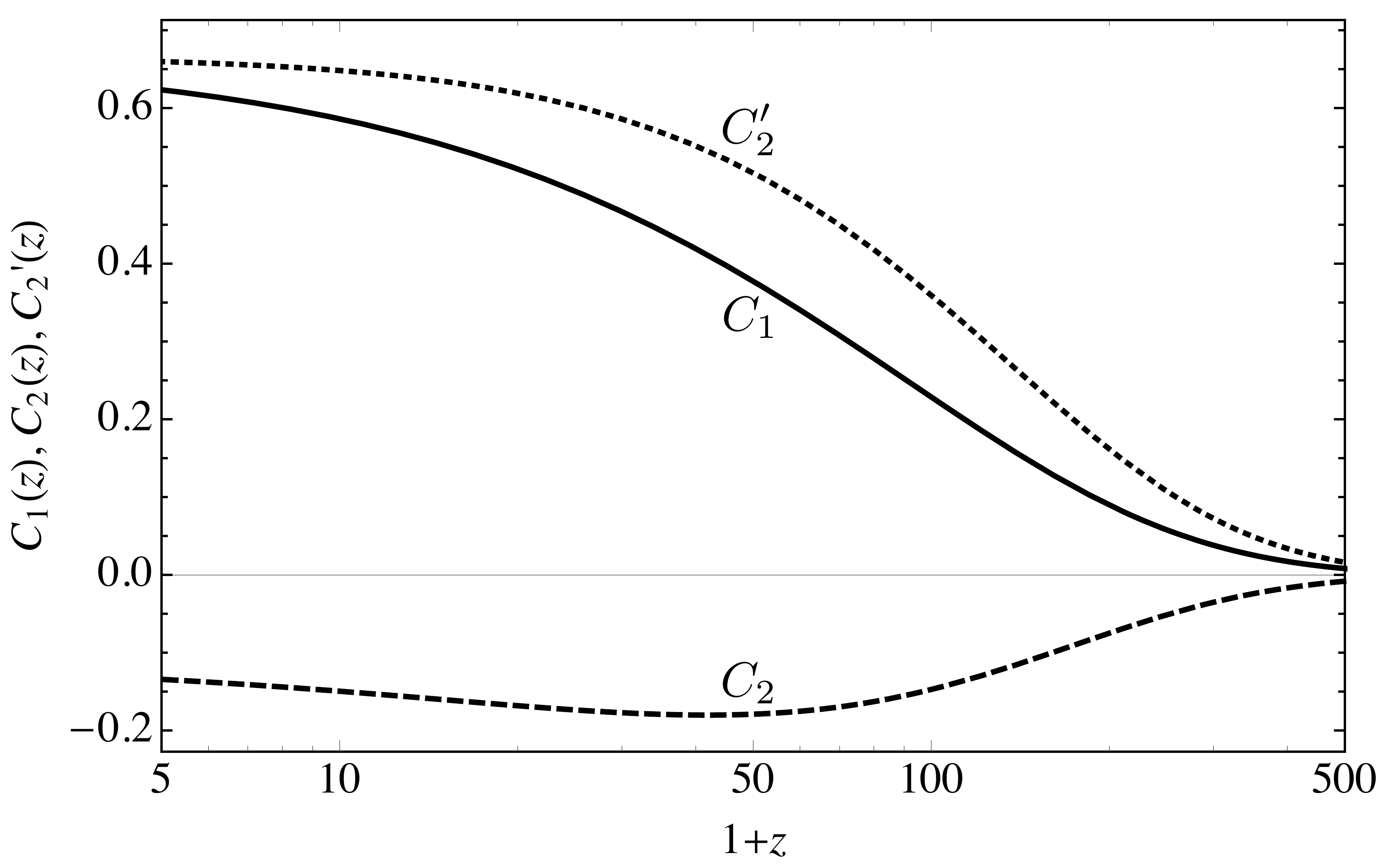}
\caption{Coefficients of the approximate decomposition of the gas-temperature fluctuations as a quadratic function of baryon density fluctuations: $\delta_{T_{\rm gas}}(\bsx, z) \approx C_1(z) \delta_b^{(1)}(\bsx, z) + C_2(z) [\delta_b^{(1)}(\bsx, z)]^2 + C_2'(z) \delta_b^{(2)}(\bsx, z)$. At high redshift, Compton heating is efficient and maintains $T_{\rm gas} = T_{\rm cmb}$, with negligible fluctuations, so $C_1 \approx C_2 \approx C_2' \approx 0$. At low redshift, the gas decouples thermally from the CMB and starts cooling down adiabatically, asymptoting towards $T_{\rm gas} \propto n_b^{2/3}$, which implies $C_1 \approx C_2' \rightarrow 2/3$ and $C_2 \rightarrow -1/9$.}
\label{fig:C1C2}
\end{figure}

The evolution of the gas temperature can be obtained from the first law of thermodynamics. Neglecting fluctuations of the CMB temperature and the effect of gravitational potentials, the full non-linear equation is \cite{1312.4948}
\barr
&& \dot{\delta}_{T_{\rm gas}} - \frac23 \dot{\delta}_b \frac{1 +
  \delta_{T_{\rm gas}}}{1 + \delta_b} =\nonumber\\
&& \Gamma_{\rm C} \left[\frac{\overline{T}_{\rm cmb} - \overline{T}_{\rm
      gas}}{\overline{T}_{\rm gas}} \delta_{x_e} -
 \left( \frac{\overline{T}_{\rm cmb}}{\overline{T}_{\rm gas}} + \delta_{x_e}\right)\delta_{T_{\rm gas}} \right],\label{eq:Tgas-nonpert}
\earr
where $\Gamma_{\rm C} \times(T_{\rm cmb} - T_{\rm gas})$ is the rate at which Thomson scattering of CMB photons by free electrons heats up the gas. Since $\Gamma_{\rm C} \propto T_{\rm cmb}^4 x_e$, the fluctuations of the gas temperature are coupled to those of the free-electron fraction $\delta_{x_e}$. In principle this equation should be solved simultaneously with the evolution of $\delta_{x_e}$, obtained by perturbing the recombination rate \cite{1312.4948}. We find that neglecting $\delta_{x_e}$ leads to errors of order $\sim 10\%$ for the linear evolution and we shall set $\delta_{x_e} \rightarrow 0$ for simplicity. With this simplification, the equation for $\delta_{T_{\rm gas}}$ to second order is 
\beq
\dot{\delta}_{T_{\rm gas}} - \frac23 \dot{\delta}_b \left(1 - \delta_b +
  \delta_{T_{\rm gas}} \right) + \frac{\overline{T}_{\rm cmb}}{\overline{T}_{\rm gas}} \Gamma_{\rm C} \delta_{T_{\rm gas}} = 0. \label{eq:dTgas-approx}
\eeq
We shall consider scales larger than the baryonic Jeans scale: $k \ll k_{\rm J} \sim 300$ Mpc$^{-1}$. On these scales baryons behave just like CDM, so their evolution equation does not depend on $T_{\rm gas}$. Given $\delta_b$, we can therefore solve for the gas-temperature fluctuations. We decompose the baryon-density fluctuation into a piece linear in the initial conditions $\delta_b^{(1)}$ and a quadratic piece $\delta_b^{(2)}$ resulting from non-linear gravitational collapse. We can then solve for the linear and quadratic parts of $\delta_{T_{\rm gas}}$:
\barr
\dot{\delta}_{T_{\rm gas}}^{(1)} + \frac{\overline{T}_{\rm cmb}}{\overline{T}_{\rm gas}} \Gamma_{\rm C} \delta_{T_{\rm gas}}^{(1)} &=& \frac23 \dot{\delta}_b^{(1)}, \label{eq:ODE_gas1}\\
\dot{\delta}_{T_{\rm gas}}^{(2)} + \frac{\overline{T}_{\rm cmb}}{\overline{T}_{\rm gas}} \Gamma_{\rm C} \delta_{T_{\rm gas}}^{(2)} &=& \frac23 \dot{\delta}_b^{(2)} + \frac23 \dot{\delta}_b^{(1)} (\delta_{T_{\rm gas}}^{(1)} - \delta_b^{(1)}). ~~\label{eq:ODE_gas2}
\earr
Our final approximation is to assume that $\delta_b^{(1)}$ is uniformly proportional to the scale factor $a$, i.e. $\delta_b^{(1)}(\bsx, a') = (a'/a) \delta_b^{(1)}(\bsx, a)$, independently of the position $\bsx$, and similarly that $\delta_b^{(2)} \propto a^2$. We then solve Eqs.~\eqref{eq:ODE_gas1} and \eqref{eq:ODE_gas2} starting at $z = 1000$ with vanishing initial conditions. The mean free-electron fraction $x_e$ required for $\Gamma_{\rm C}$ and mean gas temperature $\overline{T}_{\rm gas}$ are obtained from \textsc{HyRec} \cite{1006.1355, 1011.3758}. This allows us to obtain three coefficients $C_1(z)$, $C_2(z)$ and $C_2'(z)$ such that 
\barr
\delta_{T_{\rm gas}}^{(1)}(\bsx, z) &=& C_1(z) \delta_b^{(1)}(\bsx, z), \label{eq:dTgas_1}\\
\delta_{T_{\rm gas}}^{(2)}(\bsx, z) &=& C_2(z) [\delta_b^{(1)}(\bsx, z)]^2 + C_2'(z) \delta_b^{(2)}(\bsx, z),
\earr
which we show in Fig.~\ref{fig:C1C2}. 

The assumption that $\delta_b^{(1)} \propto a$ and $\delta_b^{(2)} \propto a^2$ is not quite correct. Indeed this assumes that baryons behave exactly like CDM. In reality, they start with different ``initial" conditions at $z \approx 1000$, after they decouple from the photon fluid shortly after cosmological recombination: their overdensity is typically significantly smaller than that of the CDM on sub-horizon scales, and their velocity field, though comparable to that of the CDM in magnitude, has a very different scale dependence (hence leading to the relative-velocity effect \cite{1005.2416}). Baryons therefore take some time to ``catch up" to the CDM, and their growth rate at early times differs from $\delta_b \propto a$, and is scale-dependent. Given that Thomson scattering maintains $T_{\rm gas}  = T_{\rm cmb}$ at $z \gtrsim 200$, regardless of the exact value of $\delta_b$, this should not be a major issue, but should be properly accounted for in a detailed analysis.

\begin{figure}
\includegraphics[width=88mm]{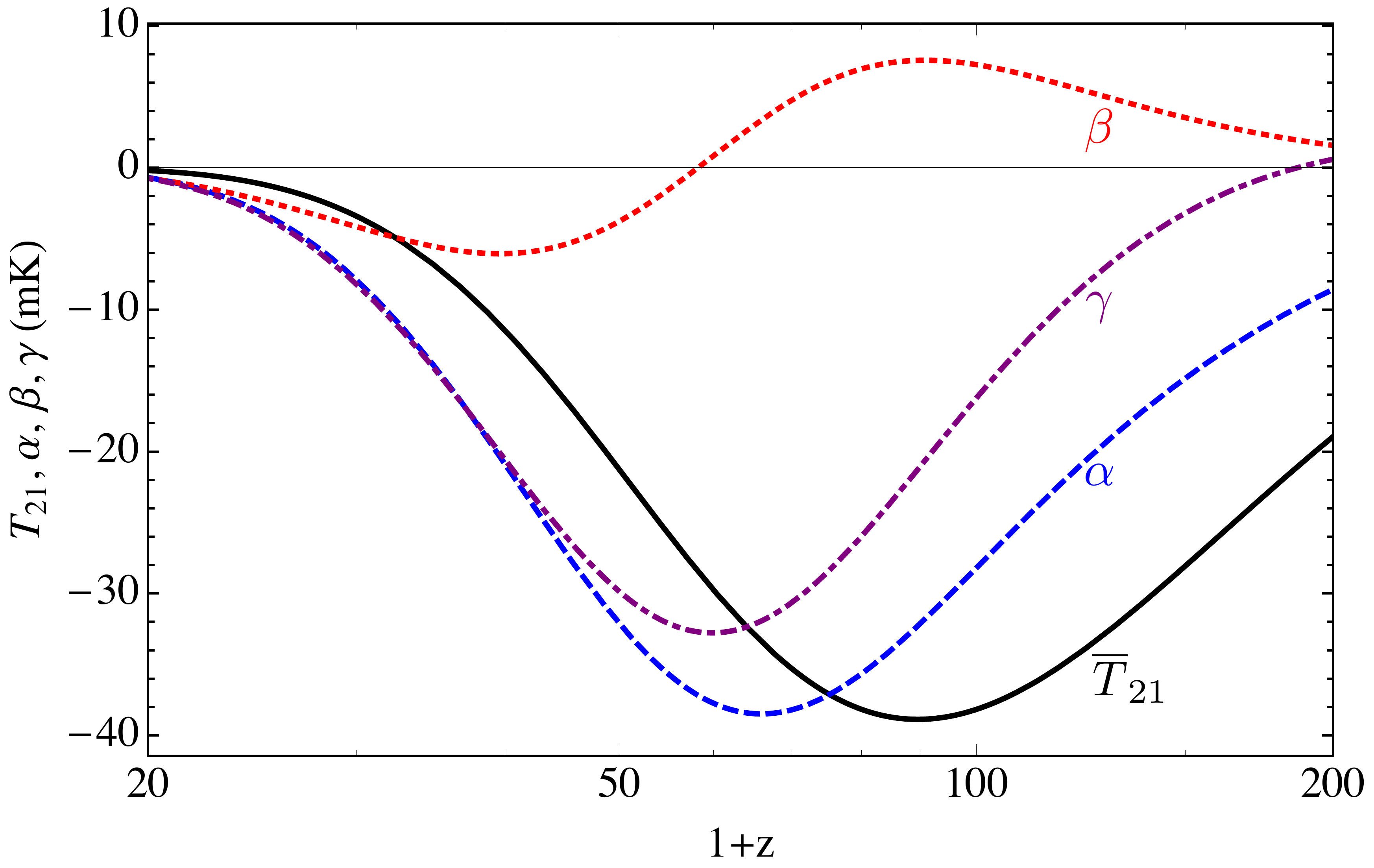}
\caption{Coefficients of the approximate decomposition of the 21-cm brightness temperature given in Eq.~\eqref{eq:T21}, in mK. In solid black we plot $\tto$, in blue dashed $\alpha=\partial T_{21}/\partial \delta_b^{(1)}$, in red dotted $\beta=\frac12 \partial^2 T_{21}/\partial \delta_b^2$, and in purple dot-dashed $\gamma = \partial T_{21}/\partial \delta_b^{(2)}$.}
\label{fig:TFs}
\end{figure}

With these caveats in mind, we substitute our approximation $\delta_{T_{\rm gas}} = C_1 \delta_b^{(1)} + C_2 [\delta_b^{(1)}]^2 + C_2' \delta_b^{(2)}$ into equation \eqref{eq:T21-full} and obtain the following simpler expression for the 21-cm brightness temperature fluctuations to second order, with which we shall work for the rest of this paper:
\barr
\delta T_{21} &\approx& \overline{T}_{21} (\delta_v^{(1)}+\delta_v^{(2)} + [\delta_v^{(1)}]^2)\nonumber\\
 &+& \alpha(z) \delta_b^{(1)} (1 + \delta_v^{(1)}) + \beta(z) [\delta_b^{(1)}]^2 + \gamma(z) \delta_b^{(2)}. \label{eq:T21}
\earr
The effective coefficients $\alpha$, $\beta$, and $\gamma$ are straightforwardly obtained from the coefficients of Eq.~\eqref{eq:T21-full} and $C_1, C_2, C_2'$ and are shown in Figure \ref{fig:TFs}. 

\subsection{Neglected sources of fluctuations}

The above analysis is only valid on subhorizon scales, and does not account for several relativistic effects. First, the gas is not at rest with respect to comoving observers. We have already accounted for the resulting perturbation to the local Hubble expansion rate due to the velocity gradient. In addition, a local velocity leads to $(i)$ a difference between the proper time in the baryon rest frame and the comoving frame, $(ii)$ a dipolar anisotropy of the CMB intensity in the baryon rest frame, and $(iii)$ an additional redshifting of the observed frequency. Gravitational potentials also affect the observed brightness temperature through: $(i)$ a time dilation, $(ii)$ a perturbation to the local expansion rate, $(iii)$ the Sachs-Wolfe and integrated Sachs-Wolfe effects\footnote{See Ref.~\cite{Raccanelli_2015} for a discussion of the ISW effect for 21-cm surveys.}, and $(iv)$ lensing by intervening structure, as is familiar from CMB studies. All these relativistic corrections are rigorously accounted for using the relativistic Boltzmann equation in Ref.~\cite{astro-ph/0702600}. They lead to fluctuations on scales comparable to the horizon at the redshift of absorption, i.e. $k \lesssim 10^{-3}$ Mpc$^{-1}$ \cite{astro-ph/0702600}. We will neglect them in this study, which is justified as we shall see that most of the signal-to-noise for PNGs comes from small scales, with $k \gg 10^{-3}$ Mpc$^{-1}$.

Redshift-space distortions are an additional source of non-linear fluctuations. The observer has only access to the total redshift $z_{\rm obs} \equiv \lambda_{\rm obs}/\lambda_{21} -1$, and will compute the angular power spectrum on slices of fixed $z_{\rm obs}$. The observed redshift is the sum of the cosmological redshift $z$ and the redshift due to the relative peculiar velocity $v_{||}$ along the line of sight: $z_{\rm obs} = z + v_{||}/c$. The observed brightness temperature at wavelength $\lambda_{\rm obs}$ is therefore
\beq
T_{21}^{\rm obs}(\lambda_{\rm obs}, \hat{n}) = \frac{T_{21}^{\rm loc}(z, \hat{n})}{1+ z_{\rm obs}},
\eeq
where the true redshift $z \equiv z_{\rm obs} - v_{||}(z, \hat{n})/c$ depends implicitly on the unknown local velocity. The angular power spectrum at fixed $z_{\rm obs}$ therefore has additional non-linear terms \cite{astro-ph/0702600,0808.1724,1109.4612}. We shall not account for those in this study but they should of course be modeled accurately when actual data is available.

Finally, Ref.~\cite{1312.4948} showed that the non-linear dependence of the 21-cm fluctuation on the local baryon density and temperature leads to enhanced large-scale fluctuations due to the relative velocity effect \cite{1005.2416}. The magnitude of the enhanced fluctuations is $\delta T_{21} \sim \beta \Delta \langle \delta_s^2 \rangle$, where $\beta$ is the coefficient of quadratic terms in the brightness-temperature fluctuations, and $\Delta \langle \delta_s^2 \rangle$ is the large-scale fluctuation of small-scale power due to the relative velocity effect. These enhanced fluctuations are most important for scales $k \lesssim 0.1$ Mpc$^{-1}$, and we will not account for them in this study, where we focus mostly on smaller scales. The relative-velocity effect also leads to a suppression of the average small-scale power, but this takes place at scales $k \gtrsim 100$ Mpc$^{-1}$, which we do not consider. 

\subsection{Flat-sky formalism}

\subsubsection{Fourier transform}

We consider a small patch on the sky, across which we can assume that the line of sight $\hat{n}$ is a constant direction. We then define the Fourier transform of the brightness temperature as
\beq
\delta T(\bsk) \equiv \int d r d^2x_{\bot} \rme^{- i \bsk \cdot \bsx} \delta T(r \hat{n}, \bs{x}_{\bot}).
\eeq
Assuming matter domination and that the baryons have caught up to the dark matter so that $\delta_b \propto a$, at linear order the peculiar velocity term is $\delta_v(\bsk) =  (\hat{k} \cdot \hat{n})^2 \delta_b(\bsk)$. The linear terms of Eq.~\eqref{eq:T21} therefore contribute a Fourier transform
\be
\delta T^{\rm lin}(\bsk) = [\alpha + \overline{T}_{21} (\hat{k} \cdot \hat{n})^2] \delta_b(\bsk). \label{eq:dT-linear}
\ee
The power spectrum of 21-cm fluctuations is therefore anisotropic: to lowest order, and defining $k_{||} \equiv \bsk \cdot \hat{n}$, 
\beq
P_{\delta T}(\bsk) = \left(\alpha + \overline{T}_{21} ~ k_{||}^2/k^2\right)^2 P_{\delta_b}(k).
\eeq
Similarly, the bispectrum $B_{\delta T}(\bsk_1, \bsk_2, \bsk_3)$ is anisotropic, and depends on the orientation of the wavenumbers with respect to the line of sight. It is defined as usual through
\barr
\langle \delta T_{21}(\bsk_1) \delta T_{21}(\bsk_2) \delta T_{21}(\bsk_3) \rangle &=& (2 \pi)^3 \delta_{\rm D}(\bsk_1 + \bsk_2 + \bsk_3) \nonumber\\
&&\times B_{\delta T}(\bsk_1, \bsk_2, \bsk_3).
\earr

\subsubsection{Harmonic transform}

Since we focus on small angular scales, we adopt a flat-sky formalism \cite{astro-ph/0001303, Bernardeau_2011}. We assume that the 21-cm temperature is observed with a finite window function $W$ in frequency. The observed temperature is therefore the convolution of the underlying temperature with $W$, which we shall denote by $W*\delta T$. We define the flat-sky harmonic transform
\be
\delta T (r, \bsl) \equiv \int_{\mathcal{A}} \frac{d^2 x_{\bot}}{r^2} \rme^{- i \bsl \cdot \bs{x}_{\bot}/r} (W*\delta T) (r \hat{n}, \bs{x}_{\bot}) ,
\ee
where $\hat{n}$ is the line of sight, assumed constant over the small survey area $\mathcal{A}$, and $\bs{x}_{\bot}$ is perpendicular to the line of sight. In terms of the Fourier modes of $\delta T$, this gives
\be
\delta T(\bsl) =  \int \frac{d^3 k}{(2 \pi)^3} \rme^{i r k_{||}} \tilde{W}(k_{||})\delta T(\bsk)  (2 \pi)^2\tilde{\delta}_{\rm D}(r \bsk_{\bot} - \bsl),
\ee
where $\tilde{W}(k_{||})$ is the Fourier transform of the window function and we have defined
\beq
\tilde{\delta}_{\rm D}(\bsl) \equiv \frac1{(2 \pi)^2}\int_{\mathcal{A}} \frac{d^2 x_{\bot}}{r^2} \rme^{i \bs{x}_{\bot} \cdot \bsl /r}.
\eeq
The function $\tilde{\delta}_{\rm D}$ peaks at the origin, with value $\tilde{\delta}_{\rm D}(\bs{0}) = f_{\rm sky}/\pi$, where $f_{\rm sky}$ is the fraction of sky subtended by the survey. If has a characteristic width $\Delta \ell \sim (f_{\rm sky})^{-1/2}$ and integrates to unity. Finally, a convolution of $\tilde{\delta}_{\rm D}$ with itself gives $\tilde{\delta}_{\rm D}$ back.

\begin{figure}[htbp!]
\centering
\hspace{0cm}
\includegraphics[width=85mm]{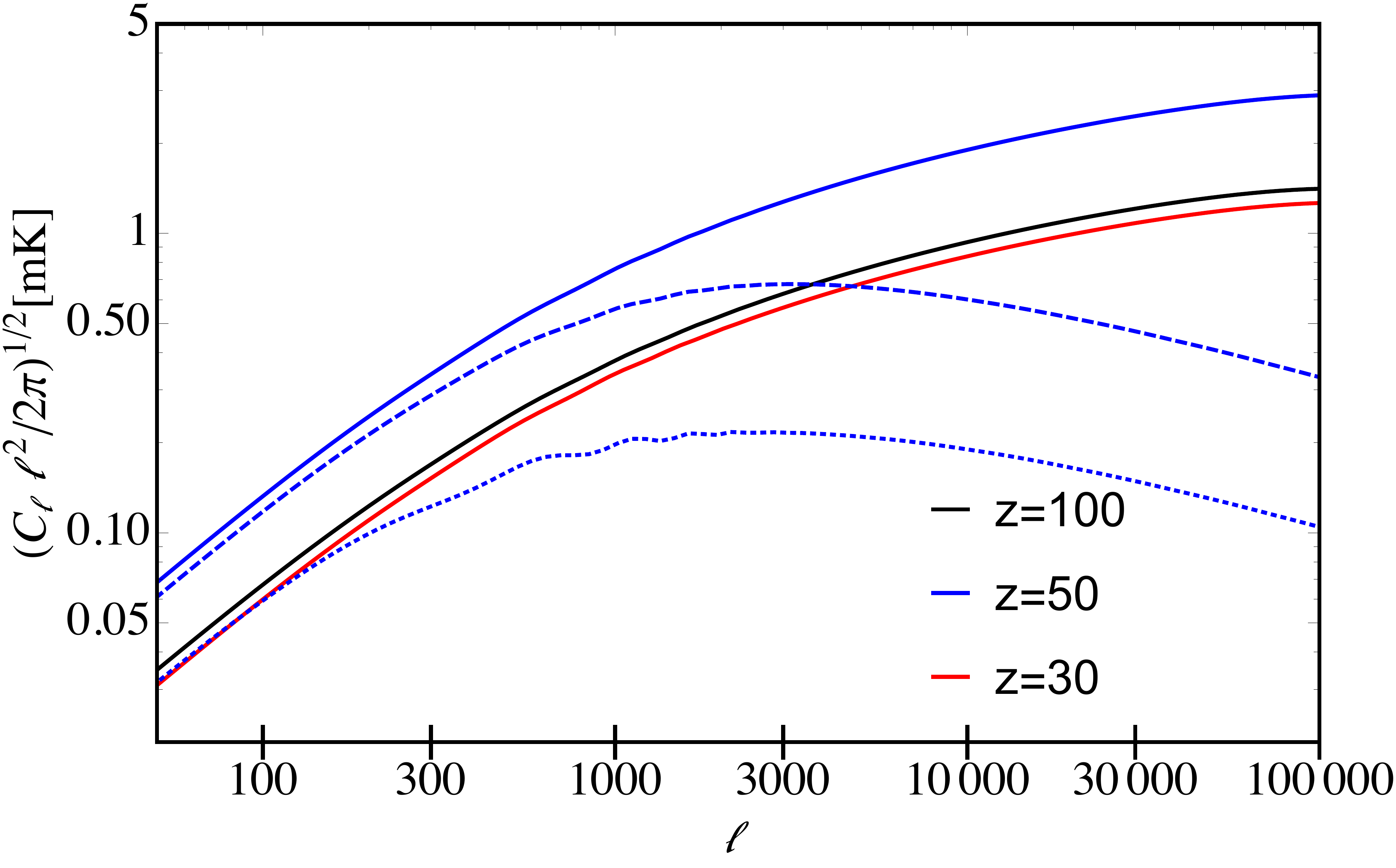}
\caption{Flat-sky power spectrum $C_{\ell}$ in the limit of infinitely narrow window function, for redshifts (top to bottom) $z=50$ (blue), $z=100$ (black) and $z=30$ (red). We also show the $C_{\ell}$ at redshift $z=50$ for a Gaussian window function of width $0.1$ MHz (blue dashed) and width of 1 MHz (blue dotted).}
\label{fig:Cls}
\end{figure}

The covariance of $\delta T(\bsl)$ at equal $r$ is given by
\barr
\langle \delta T(\bsl) \delta T^*(\bsl') \rangle 
&=& \int d^2 k_{\bot} (2 \pi)^2\tilde{\delta}_{\rm D}(r \bsk_{\bot} - \bsl)\tilde{\delta}_{\rm D}^*(r \bsk_{\bot} - \bsl') \nonumber\\
&& ~~ \times \int \frac{dk_{||}}{2 \pi} |\tilde{W}|^2(k_{||}) P_{\delta T}\left(k_{||}, \bsk_{\bot}\right).
\earr
For $\ell\gg (f_{\rm sky})^{-1/2}$, we may approximate $\bsk_{\bot} \approx \bsl/r$ in the inner integral. Carrying out the outer integral, we arrive at
\be
\langle \delta T(\bsl) \delta T^*(\bsl') \rangle \approx (2 \pi)^2 \tilde{\delta}_{\rm D}(\bsl' - \bsl) C_{\ell},
\ee
where \cite{Bernardeau_2011}
\be
C_{\ell} \equiv \frac1{r^2} \int \frac{d k_{||}}{2 \pi} |\tilde{W}|^2(k_{||}) P_{\delta T}\left(k_{||}, \bsl/r\right).\label{eq:Cl}
\ee
We show the flat-sky power spectrum $C_{\ell}$ computed with different widths of the window function and for several redshift slices in Fig.~\ref{fig:Cls}.

Similarly, the three-point function of $\delta T(\bsl)$ defines our flat-sky bispectrum:
\be
\langle \delta T(\bsl_1) \delta T(\bsl_2)  \delta T(\bsl_3) \rangle = (2 \pi)^2 \tilde{\delta}_{\rm D}(\bsl_1 + \bs{\ell}_2 + \bsl_3) B_{\ell_1 \ell_2 \ell_3},
\label{eq:bispharm}
\ee
with \cite{Bernardeau_2011}
\barr
B_{\ell_1 \ell_2 \ell_2}  \equiv \int \frac{d k_{1||} d k_{2||}}{(2 \pi)^2 r^4} \tilde{W}(k_{1||}) \tilde{W}(k_{2||}) \tilde{W}(-k_{1||} - k_{2||}) \nonumber\\
 B_{\delta T}\left(k_{1 ||}, \bsl_1/r; k_{2 ||}, \bsl_2/r\right), ~~~~~~ \label{eq:Bflat}
\earr
where we have dropped the dependence on $\bsk_3$ in the $k$-space bispectrum since it is fixed by the triangle condition given $\bsk_1 = k_{1||} \hat{n} + \bsl_1/r$ and $\bsk_2 = k_{2||} \hat{n} + \bsl_2/r$. Note that we do not use the Limber approximation and perform the full integrals over $k_{||}$'s.

We now describe and compute the different contribution to $B_{\ell_1 \ell_2 \ell_3}$.

\section{Bispectrum of 21-cm fluctuations}
\label{sec:nong}

The bispectrum gets contributions from primordial non-gaussianities, which we would like to extract from the data, but also from secondary non-gaussianities, arising from the non-linear relation between the observable and the initial conditions, even if the latter are perfectly gaussian.

We only consider multipoles $\ell \gtrsim 100$, which correspond to wavenumbers $k \gtrsim 0.01$ Mpc$^{-1}$, as most of the signal-to-noise ratio for bispectrum measurements is expected to come from small-scale modes. We therefore neglect the contributions of relativistic terms to the bispectrum, in particular the ISW-lensing bispectrum, which is the dominant secondary bispectrum for CMB anisotropies \cite{0905.4732,astro-ph/9811251,astro-ph/0612571,1303.5084}.

\subsection{Primordial non-gaussianities}\label{sec:PNG}

The contribution of PNG to the bispectrum of 21-cm fluctuations can be obtained to lowest order by only considering the linear terms in Eq.~\eqref{eq:T21}, and assuming that they are linearly related to the primordial curvature fluctuations. The Fourier transform of the linear terms is given in Eq.~\eqref{eq:dT-linear}. We define $M(k, z) \equiv \delta_b(\bsk, z)/\Phi(\bsk)$, where $\Phi = (3/5) \zeta$ is Bardeen's gravitational potential. The bispectrum of brightness-temperature fluctuations gets a contribution  
\barr
B_{\delta T}^{\rm prim}(\bsk_1, \bsk_2, \bsk_3) &=& \prod_{i = 1}^3 (\alpha + \overline{T}_{21} \mu_i^2) M(k_i)  \nonumber\\
&&~~~~ \times B_{\Phi}(k_1, k_2, k_3) \label{eq:Bprim}
\earr
from primordial non-gaussianities, where $\mu_i \equiv (\bsk_i \cdot \hat{n})/k_i$. 

We now review the different shapes of the initial potential bispectrum $B_{\Phi}(k_1, k_2, k_3)$ that we will consider in this paper (see e.~g.~Ref.~\cite{1502.01592} for a larger variety of shapes).

\subsubsection{Local}

The simplest form of PNG is of the local type, where the primordial potential $\Phi$ is a local non-linear function of a gaussian field $\phi$: 
\beq
\Phi(\bs x) = \phi(\bs x) +\f^{\mathrm{local}}  \left ( \phi^2(\bsx) -\VEV{\phi}^2\right).
\eeq
This implies a non-vanishing bispectrum for $\Phi$, given to lowest order by 
\barr
B_{\Phi}^{\rm local}(k_1, k_2, k_3) = 2 f_{\rm NL}^{\rm local} \left[P_\Phi(k_1)P_\Phi(k_2) + 2 \textrm{ perm.}\right].~~~~~
\earr
This form of the bispectrum peaks in the squeezed configuration ($k_1 \ll k_2\sim k_3$ and permutations). 

Local-type PNG typically arises in multi-field inflation models, such as the curvaton model or modulated reheating \cite{Byrnes_2010}.

\subsubsection{Equilateral}

PNGs of the equilateral type arise when there are non-standard kinetic terms in the inflation Lagrangian, which are included in the so-called $P(X)$ models of inflation \cite{hep-th/0605045}, concrete examples of which are k-inflation \cite{hep-th/9904075,hep-th/9904176} and Dirac-Born-Infield inflation \cite{hep-th/0310221,hep-th/0404084}. In these models the effective sound speed $c_s$ can be very different from the speed of light ($c = 1$), and the non-gaussianity parameter is related to this departure via $\f^{\mathrm {equil}} = -(35/108)(c_s^{-2}-1)$ \cite{hep-th/0605045}.

This shape peaks when the three modes cross the horizon at the same time, and hence $k_1\sim k_2 \sim k_3$. A good
template for it is \cite{astro-ph/0509029},
\barr
B_{\Phi}^{\mathrm {equil}}(k_1, k_2, k_3) = 6 f_{\rm NL}^{\rm equil} A_\Phi^2 \Bigg{\{} -\left[\dfrac 1 {(k_1 k _2)^{4-n_s}} + \textrm{ 2 perm.} \right]\nonumber\\
 - \dfrac 2 {\left(k_1 k_2 k_3 \right)^{2 \frac {4-n_s}{3}}} + \left [ \dfrac 1 {(k_1k_2^2 k_3^3)^{\frac{4-n_s}{3}}} +\textrm{ 5 perm.} \right]   \Bigg{\}},~~~~~
\earr
where $A_\Phi$ is the normalization of the power spectrum of $\Phi$: $P_{\Phi}(k) = A_\Phi/ k^{4 - n_s}$. 

\subsubsection{Orthogonal}

The ``orthogonal" shape of PNG was defined in Ref.~\cite{0905.3746} to be orthogonal to the equilateral shape for the scalar product $B^a \cdot B^b \equiv \sum_{k_1, k_2, k_3} B^a_{k_1, k_2, k_3} B^b_{k_1, k_2, k_3}/[P_{\Phi}(k_1) P_{\Phi}(k_2) P_{\Phi}(k_3)]$. Its form is
\barr
B_{\Phi}^{\mathrm {ortho}}(k_1, k_2,k_3) = 6 f_{\rm NL}^{\rm ortho} A_\Phi^2 \Bigg{\{} \left[\dfrac {-3} {(k_1 k _2)^{4-n_s}} + \textrm{ 2 perm.} \right]\nonumber\\
 - \dfrac 8 {\left(k_1 k_2 k_3 \right)^{2 \frac {4-n_s}{3}}} + \left [ \dfrac 3 {(k_1k_2^2 k_3^3)^{\frac{4-n_s}{3}}} +\textrm{ 5 perm.} \right]   \Bigg{\}}.~~~~~
\earr

The models of Galileon inflation \cite{1009.2497} and ghost inflation \cite{hep-th/0312100} predict very high values of $\f^{\rm ortho}$. In general, in terms of the Lagrangian for the Goldstone boson $\pi$ during inflation, both equilateral and orthogonal shapes arise from cubic kinetic interactions, and the $\f$s are linearly related to the coefficients of the $\dot\pi^3$ and $\dot\pi(\partial\pi)^2$ terms \cite{0905.3746}.

It is interesting to also mention the folded form of non-gaussianity \cite{0901.4044}, where the shape of the bispectrum peaks at
flattened (folded) triangles ($k_1 = k_2 = k_3/2$ and permutations). Initial conditions different from the standard Bunch-Davies vacuum would give rise
to this kind of PNG \cite{1012.3392}. It can be expressed as a combination of the two above, as $
B^{\mathrm {folded}} = \left(B^{\mathrm {equil}} - B^{\mathrm {ortho}}\right )/2$.

\subsubsection{Directional dependence}

In some models where inflation is driven by a gauge vector field \cite{1202.2847} or in solid inflation \cite{1210.0569} there is an additional form
of PNG, that induces an extra dependence in the angle between the $\bs{k}_i$ vectors.
In this case the bispectrum can be decomposed in Legendre polynomials \cite{1302.3056}, where each component would be
\barr
B_{\Phi}^{(J)}(k_1, k_2, k_3) &=& f_{\rm NL}^{(J)} \Big{[}P_{\Phi}(k_1) P_{\Phi}(k_2) \mathcal{P}_J(\cos \theta_{12}) \nonumber\\
&& ~~~~~~~~~~ + 2 \textrm{ perm.}\Big{]},
\earr
where $\mathcal{P}_J$ is the Legendre polynomial of order $J$, and $\theta_{12}$ is the angle between $\bsk_1$ and $\bsk_2$, whose cosine can be expressed as $\cos\theta_{12} = (k_3^2 - k_1^2-k_2^2)/2k_1k_2$. We consider $J=1$, 2 and 3.

\subsection{Secondary non-gaussianities} \label{sec:secondaries}

\subsubsection{Non-linear gravitational collapse}

The growth of overdensities by gravitational collapse is a fundamentally non-linear process, leading to a non-vanishing 3-point function, even when starting from perfectly gaussian initial conditions. The resulting bispectrum can be computed from second-order perturbation theory (see e.g.~Ref.~\cite{astro-ph/0112551}). The correlation of two linear perturbations with a second-order density perturbation or normalized velocity divergence ($\theta \equiv - \bs{\nabla}\cdot \bs{v}/H$) takes the form  

\barr
\langle \delta^{(1)}(\bsk_1) \delta^{(1)}(\bsk_2) \delta^{(2)}(\bsk_3) \rangle' &=& 2 F(\bsk_1, \bsk_2) P_1 P_2, \\
 \langle \delta^{(1)}(\bsk_1) \delta^{(1)}(\bsk_2) \theta^{(2)}(\bsk_3) \rangle' &=& 2 G(\bsk_1, \bsk_2)P_1 P_2,
\earr
where $\langle ... \rangle'$ is the three-point function divided by $(2 \pi)^3 \delta_{\rm D}(\bsk_1 + \bsk_2 + \bsk_3)$, and $P_i \equiv P_{\delta}(\bsk_i)$ is the power spectrum of the linear overdensity.
The mode-coupling kernels $F(\bsk_1, \bsk_2)$ and $G(\bsk_1, \bsk_2)$ are both of the form \cite{Fry, astro-ph/0112551}\footnote{These coupling kernels are derived in the sub-horizon limit. Since we are mostly interested in small scales we shall not concern ourselves with subtle issues regarding the squeezed limit of the gravitational bispectrum on horizon scales \cite{1504.00351, dePutter_2015, Bartolo_2015}.}
\beq
K(\bsk_1, \bsk_2) = c_1 +c_2~ \hat{k}_1 \cdot \hat{k}_2 \left( \dfrac {k_1} {k_2} +\dfrac {k_2} {k_1} \right) + c_3 (\hat{k}_1 \cdot \hat{k}_2)^2. \label{eq:F-kernel}
\eeq
For a CDM-only universe, $(c_1, c_2, c_3) = (\frac57, \frac12, \frac27)$ for $F$ and $(\frac37, \frac12, \frac47)$ for $G$. In reality, however, baryons start clustering after recombination, while the CDM overdensities have already been growing since their scales entered the horizon. Their density and velocity fields at recombination are therefore very different and the subsequent growth factor of matter fluctuations is therefore not just $D(a) \propto a$. It moreover has a scale dependence, as baryons, though they start with effectively zero overdensity at recombination ($\delta_b \ll \delta_c$ on sub-horizon scales), have a velocity comparable to that of the CDM, but with a different scale dependence. Different wavenumbers therefore grow at slightly different rates. The coefficients $c_i$ in Eq.~\eqref{eq:F-kernel} are therefore in reality weakly dependent on redshift and, perhaps to a lesser extent, on scale \cite{YAH_Jeong}. We shall ignore these complications here and take their standard values.

Assuming $\delta_b = \delta$ and using $\delta_v(\bsk) = \mu^2 \theta(\bsk)$ and $\delta_v^{(1)}(\bsk) = \mu^2 \delta_b^{(1)}(\bsk)$ [note that this last relation only holds for the first-order perturbations], the bispectrum of 21-cm fluctuations due to gravitational collapse is straightforwardly obtained from Eq.~\eqref{eq:T21}:
\barr
B_{\delta T}^{\rm grav}(\bsk_1, \bsk_2, \bsk_3) = 2 (\alpha +  \overline{T}_{21} \mu_1^2)(\alpha + \overline{T}_{21} \mu_2^2) \nonumber\\
\times \left(\gamma F(\bsk_1, \bsk_2) + \overline{T}_{21} (\mu_1 + \mu_2)^2 G(\bsk_1, \bsk_2) \right) P_1 P_2 \nonumber\\
+ 2 \textrm{ perm.} \label{eq:secondary-grav}
\earr

\subsubsection{Non-linear relation between brightness temperature and baryon density}

The relationship between the 21-cm brightness temperature and the underlying density and velocity field is fundamentally non-linear, due to $(i)$ the non-linear dependence of the optical depth on the local peculiar velocity gradient ($\tau\propto 1/(1-\delta_v)$), $(ii)$ the non-linear dependence of the spin temperature on the baryon density and temperature, and $(iii)$ the non-linear dependence of the gas temperature on the baryon density. Therefore even for a perfectly gaussian underlying density field, this non-linear mapping leads to  a non-vanishing bispectrum. 

This contribution to the bispectrum can be obtained from the following three-point functions: 
\barr
\langle \delta_b(\bsk_1) \delta_b (\bsk_2) [\delta_b^2](\bsk_3) \rangle' &=& 2 P_1 P_2,\\
 \langle \delta_b(\bsk_1) \delta_b (\bsk_2) [\delta_b \delta_v](\bsk_3) \rangle' &=& (\mu_1^2 + \mu_2^2) P_1 P_2,\\
\langle \delta_b(\bsk_1) \delta_b (\bsk_2) [\delta_v^2](\bsk_3) \rangle'&=& 2 \mu_1^2 \mu_2^2 P_1 P_2,
\earr
where the superscript $(1)$ is implicit in all the fluctuations. Using Eq.~\eqref{eq:T21}, the explicit expression for the bispectrum arising from the non-linearity of $\delta T_{21}$ as a tracer is then
\barr
B_{\delta T}^{\rm nl}(\bsk_1, \bsk_2, \bsk_3) = (\alpha + \overline{T}_{21} \mu_1^2)(\alpha + \overline{T}_{21} \mu_2^2) \nonumber\\
\times \left(2 \beta + \alpha (\mu_1^2 + \mu_2^2) + 2 \overline{T}_{21} \mu_1^2 \mu_2^2 \right) P_1 P_2\nonumber\\
+ 2 \textrm{ perm.} \label{eq:secondary-nl}
\earr

The total secondary bispectrum is obtained by summing Eqs.~\eqref{eq:secondary-grav} and \eqref{eq:secondary-nl}. Note that the bispectrum arising from Eq.~\eqref{eq:secondary-grav} requires the kernels $F$ and $G$ to be non-zero, whereas the bispectrum from \eqref{eq:secondary-nl} does not.

\subsection{Numerical evaluation and comparison}

Inserting the Fourier-space primordial bispectra Eq.~\eqref{eq:Bprim} and secondary bispectra Eqs.~\eqref{eq:secondary-grav} and \eqref{eq:secondary-nl} into Eq.~\eqref{eq:Bflat}, we obtain the harmonic-space bispectra in the flat-sky limit. 

\begin{figure}[htbp!]
\centering
\includegraphics[width=85mm]{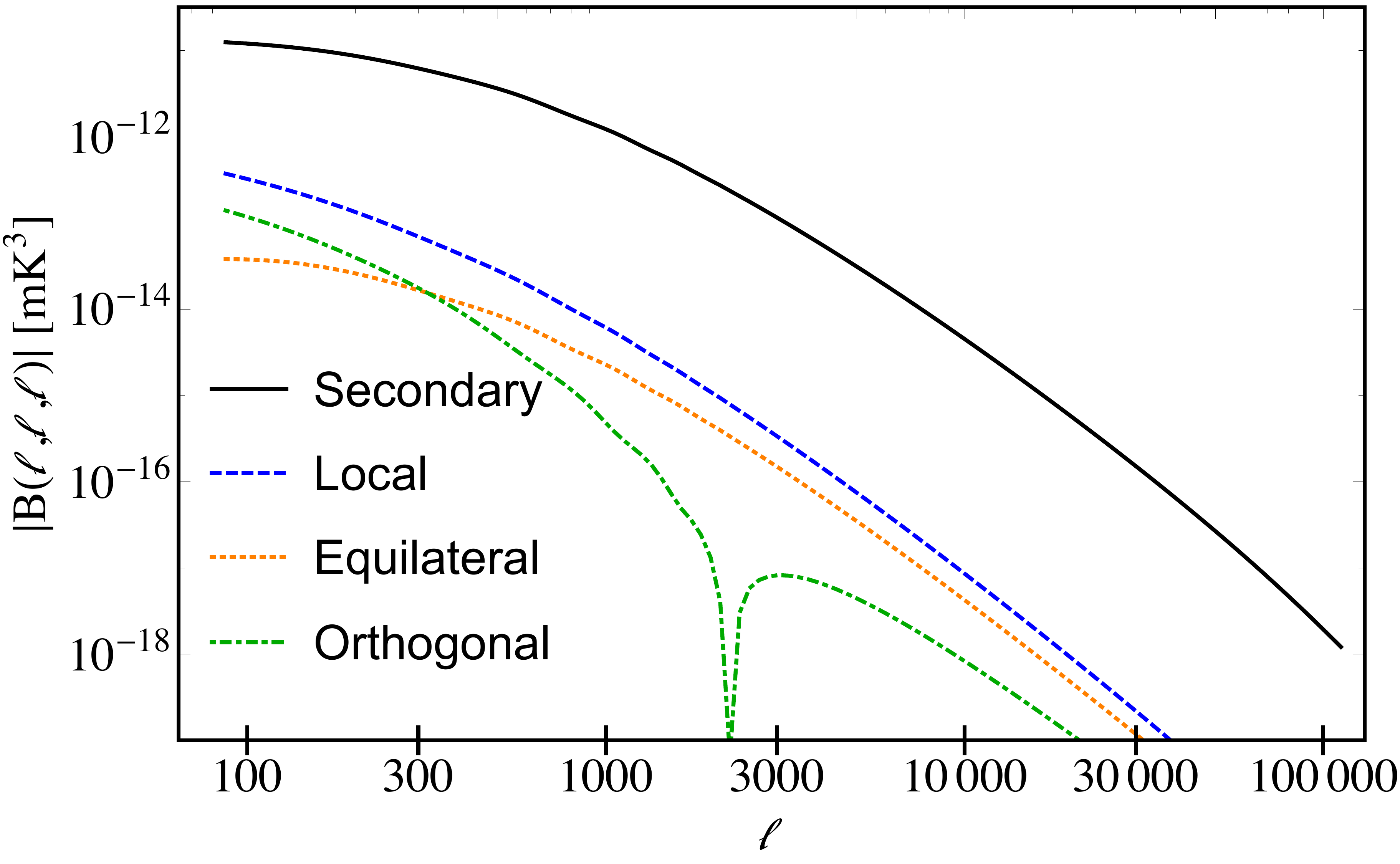}
\includegraphics[width=85mm]{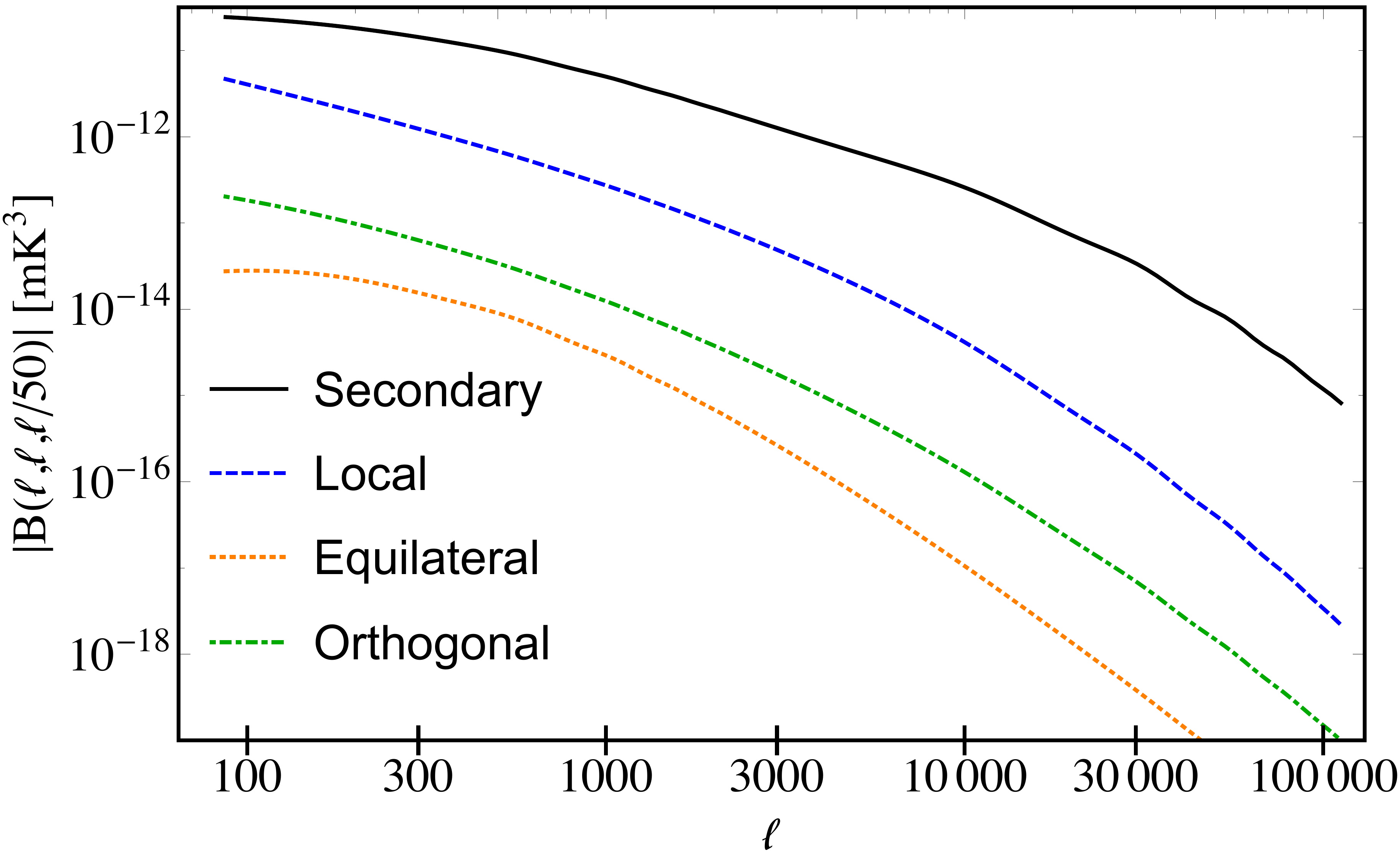}
\caption{Bispectra of 21-cm brightness-temperature fluctuations resulting from secondary non-gaussianities and different shapes of primordial non-gaussianity, with $f_{\rm NL} = 1$, at $z = 50$. The top panel shows the bispectra for equilateral triangles ($\ell \equiv  \ell_1=\ell_2=\ell_3$). The bottom panel shows the bispectra for squeezed triangles ($\ell \equiv \ell_1=\ell_2\gg \ell_3=\ell/50$). In dashed blue we plot local, in dotted orange equilateral and in dash-dotted green orthogonal non-gaussianity. In solid black we plot the secondary bispectrum. The bispectra are computed in the flat-sky approximation for an infinitesimally narrow redshift slice.}
\label{fig:bisp}
\end{figure}

We show the total secondary bispectrum in Fig.~\ref{fig:bisp}, along with the bispectra resulting from local, equilateral and orthogonal PNGs. As found by previous authors \cite{astro-ph/0611126}, we find that the secondary bispectrum is typically at least two orders of magnitude larger than the bispectrum due to PNGs for $f_{\rm NL} = 1$. This order-of-magnitude difference can be understood quite simply: the ratio of secondary to primary bispectra is of order 
\beq
\frac{B^{\rm sec}}{B^{\rm prim}} \sim \frac{\langle\delta \delta \delta \delta\rangle}{\langle \delta \delta \delta f_{\rm NL} \Phi \rangle } \sim \frac{\delta(z)}{f_{\rm NL} \Phi}.
\eeq
We know that $\delta(z = 0) \sim 1$ at the non-linear scale $k_{\rm NL}(z = 0) \approx 0.1$ Mpc$^{-1}$. Scaling back to $z = 100$ gives $\delta(z = 100) \sim 10^{-2}$ at $k \sim 0.1$, with an amplitude increasing logarithmically with wavenumber. The primordial gravitational potential is nearly scale-independent and of order $\Phi \sim 3 \times 10^{-5}$. We therefore obtain $B^{\rm sec}/B^{\rm prim} \sim$ few $\times$ 100 for $f_{\rm NL} = 1$, consistent with our more detailed calculation. 

Note that our estimates for both the primary and the secondary bispectra neglected higher-order terms, for example terms of order $\langle \delta^2 \delta^1 \delta^3 \rangle$ in the secondary bispectrum. These terms are suppressed by an additional factor of order $ \delta^2 \sim 10^{-4}$ at $z \approx 100$, and are therefore comparable to the primary bispectrum only if $f_{\rm NL} \sim$ few times $10^{-2}$. We will not consider them in this study, but they should be accounted for in a final data analysis aiming for a few percent uncertainty in $\f$. 

In practice, we carry out the integrals up to some maximum multipole $\ell_{\max}$ corresponding to the resolution of the observations.

\section{Fisher analysis}

\subsection{Bias due to secondary non-gaussianities}
\label{sec:bias}

Assuming a single type of primordial non-gaussianity with bispectrum $B_{\ell_1 \ell_2 \ell_3} = f_{\rm NL} b^{\rm prim}_{\ell_1 \ell_2 \ell_3}$, if secondary non-gaussianities were negligible the minimum-variance cubic estimator for $f_{\rm NL}$ from a single redshift $z$ would be \cite{1211.3417,astro-ph/0408455,1104.0930}
\barr
\hat{f}_{\rm NL} = \frac{(b^{\rm prim}, B^{\rm obs})_z}{(b^{\rm prim}, b^{\rm prim})_z},\label{eq:est-naive}
\earr
where we defined 
\beq
B^{\rm obs}_{\bsl_1, \bsl_2, \bsl_3} \equiv \frac1{4 \pi f_{\rm sky}} \delta T(\bsl_1) \delta T(\bsl_2) \delta T(\bsl_3),
\eeq
and the scalar product $(~, ~)_z$ is constructed as
\barr
(B^i, B^j)_z  &\equiv& 4 \pi f_{\rm sky} \iiint \limits_{\ell_1 \geq \ell_2 \geq \ell_3} \frac{d^2 \ell_1 d^2 \ell_2 d^2 \ell_3}{(2 \pi)^4}  ~\delta_{\rm D}(\bsl_1 + \bsl_2 + \bsl_3) \nonumber\\
&&~~~~~~~~~~~~~~\times\frac{B^i_{\ell_1 \ell_2 \ell_3}B^j_{\ell_1 \ell_2 \ell_3} }{C_{\ell_1}^{\rm tot} C_{\ell_2}^{\rm tot} C_{\ell_3}^{\rm tot}}. \label{eq:scal-prod}
\earr
In this equation $C^{\rm tot}_{\ell}$ is the total variance of $\delta T(\bsl)$, due to cosmic variance and all other sources of noise, including the instrument and foregrounds. We assume throughout that the noise can be approximately computed neglecting non-gaussian contributions to $\delta T(\bsl)$. In practice, all our results will be quoted in the cosmic-variance limit, i.e.~for $C_{\ell}^{\rm tot} = C_{\ell}$ given in Eq.~\eqref{eq:Cl}
\footnote{Note that in order to have tenth-of-arcminute resolution at redshift $z=100$ one would need a baseline $D\gtrsim 350$ km. In order to reach cosmic variance limit at $z = 50$ and for resolution of one arcminute, the parameters of the interferometer would have to be really optimistic, with complete coverage $f_{\rm cover}=1$, a baseline of order the diameter of the moon $D=3500$ km, and a time of observation of 2 years.}.

We saw in the previous Section that the bispectrum resulting from secondary non-gaussianities is much larger than the one arising from PNGs, typically by two to three orders of magnitude for $f_{\rm NL} = 1$. Using the estimator, Eq.~\eqref{eq:est-naive}, would therefore lead to a bias 
\beq
\Delta f_{\rm NL} = \frac{(b^{\rm prim}, B^{\rm sec})_z}{(b^{\rm prim}, b^{\rm prim})_z} \equiv c_{\rm prim, sec} \sqrt{\frac{(B^{\rm sec}, B^{\rm sec})_z}{(b^{\rm prim}, b^{\rm prim})_z}}, \label{eq:fNL-bias}
\eeq
where $c_{\rm prim, \rm sec} \in [-1, 1]$ quantifies the shape overlap or degeneracy of the primordial and secondary bispectra [geometrically, $c_{\rm prim, \rm sec}$ is the cosine of the angle between the two bispectra, for the scalar product \eqref{eq:scal-prod}]. The shapes of primordial and secondary non-gaussianity being different, one may hope that their overlap is small \cite{astro-ph/0610257}\footnote{Ref.~\cite{astro-ph/0610257} treat the secondary non-gaussianity as a source of noise instead of a bias, which is inappropriate.}. However, assuming a cosmic-variance-limited experiment with an infinitely narrow window function and a resolution of $\sim 0.1'$ (corresponding to $\ell_{\max} = 10^5$), we find that $c_{\rm loc, sec} = 0.89$,  $c_{\rm equi, sec} = 0.79$, and $c_{\rm ortho, sec} = -0.83$. The unsubtracted secondary bispectrum would therefore lead to large biases $\Delta f_{\rm NL}^{\rm loc} = 870 $, $\Delta f_{\rm NL}^{\rm equi} = 3900$, and $\Delta f_{\rm NL}^{\rm ortho} = -3900$.
For maximum resolution of $1'$ (corresponding now to $\ell_{\max} = 10^4$), the values of the degeneracy coefficients would be $c_{\rm loc, sec} = 0.80$,  $c_{\rm equi, sec} = 0.89$, and $c_{\rm ortho, sec} = -0.88$, which in turn would make the biases $\Delta f_{\rm NL}^{\rm loc} = 420 $, $\Delta f_{\rm NL}^{\rm equi} = 2400$, and $\Delta f_{\rm NL}^{\rm ortho} = -2100$. 

Such a strong degeneracy may seem surprising at first, given the large number of triangles on which the scalar product depends. However, because the bispectra are essentially smooth featureless functions of $\ell$ for small angular scales, they can have significant overlap in the sense defined in Eq.~\eqref{eq:fNL-bias}\footnote{Consider for instance the 1-dimensional scalar product $\langle F.G \rangle = \int_{\ell_{\rm min}}^{\ell_{\max}} F(\ell) G(\ell) d \ell$, with $\ell_{\min} \ll \ell_{\max}$. If $F(\ell) \propto \ell^{\alpha}$ and $G(\ell) \propto \ell^{\beta}$, then their degeneracy coefficient is approximately $c = \sqrt{1 + 2 x}/(1 + x)$, where $x \equiv (\beta - \alpha)/(2 \alpha + 1)$. This is greater than 0.4 for $x \leq 10$.}. The equilateral and orthogonal-type bispectra have more complex shapes in $k$ and $\ell$-space than the local type, which is why their overlap with the secondary bispectrum decreases with increasing $\ell_{\max}$ while it increases for the latter.

In the next section we describe how to deal with these degeneracies.

\subsection{Estimators for a single redshift slice}

One could in principle try and model the secondary bispectrum from first principles and subtract the resulting bias $\Delta f_{\rm NL}$ from the estimated PNG amplitude. This strategy is the one adopted for the bispectrum of CMB anisotropies, where the main contaminant is the ISW-lensing bispectrum. Given the now well-measured cosmological parameters, the latter can indeed be modeled to sufficient accuracy, i.e.~with an error smaller than the statistical uncertainty in $f_{\rm NL}$ \cite{astro-ph/0001303, Hanson_09, Lewis_11, 1502.01592}. In the case of 21-cm fluctuations, however, even percent-level residuals in the modeled secondary bispectrum would lead to biases of order $\Delta f_{\rm NL} \sim 10$, significantly larger than the statistical errors one may hope to achieve. Reaching sub-percent accuracy would require, first, a very careful treatment of subtle microphysical processes affecting the population of the hyperfine levels \cite{Hirata_07}. In addition, it will be limited by the accuracy of cosmological parameters.

The amplitudes of all the secondary bispectra depend on the coefficients $c_i$ in Eq.~\eqref{eq:F-kernel} for the kernels $F$ and $G$. Although we use the values derived for a matter-dominated, CDM-only universe in our analysis, we assume that the exact values could be computed exactly should one want to do so \cite{YAH_Jeong}. On the other hand, the four parameters $A_i \equiv \overline{T}_{21}, \alpha, \beta, \gamma$ in Eq.~\eqref{eq:T21} depend on the detailed microphysics of the hyperfine transition. For now we assume that they can be modeled up to subpercent accuracy and denote their best estimates by $A_i^0$. 

Our model for the bispectrum is 
\beq
B_{\ell_1 \ell_2 \ell_3} = B_{\ell_1 \ell_2 \ell_3}^{\rm sec, 0} + f_{\rm NL} ~b^{\rm prim}_{\ell_1 \ell_2 \ell_3} + \sum_{i = 1}^4 f_i ~b^i_{\ell_1 \ell_2 \ell_3}, 
\eeq
where $B_{\ell_1 \ell_2 \ell_3}^{\rm sec, 0}$ is the best estimate for the secondary bispectrum obtained with the $A_i^0$, $f_i \equiv \Delta A_i$ are the unknown residuals of the four coefficients and $b^i_{\ell_1 \ell_2 \ell_3} \equiv \partial B_{\ell_1 \ell_2 \ell_3}^{\rm sec}/\partial A_i$. 
To make the notation more compact we denote\footnote{Note that $f_{\rm NL}$ and $f_i$ do not have the same dimensions, but this does not affect the analysis.} $f_0 \equiv f_{\rm NL}$ and $b^0_{\ell_1 \ell_2 \ell_3} \equiv b^{\rm prim}_{\ell_1 \ell_2 \ell_3}$, and recall that we search for one type of PNG at a time.
Notice that we disregard higher order correction terms, proportional to $\Delta \alpha ^2$ and $\Delta \bar T_{21}^2$, since we will be able to model those two parameters to a precision better than 0.3\%, as shown in Tab.~\ref{tab:Cov}, which would mean that the bias in the non-gaussianity amplitude is of order $\Delta f_{\rm NL} \sim 10^3 (3 \times 10^{-3})^2 \lesssim 10^{-2}$.

We fit simultaneously for the amplitude of the PNG and for the residual coefficients of the secondary bispectrum. We treat the latter as nuisance parameters over which we will marginalize. In geometrical terms, we construct an estimator for the PNG by projecting the observed bispectrum on the component of the primordial bispectrum orthogonal to all shapes of secondary non-gaussianity.

The minimum-variance cubic estimators for the parameters $f_i$ are given by \cite{1211.3417}
\barr
&&\hat{f}_i \equiv \sum_j(F^{-1})_{ij} (b^j, B^{\rm obs} - B^{\rm sec, 0})_z, 
\earr
where $F^{-1}$ is the inverse of the Fisher matrix $F$ whose components are 
\beq
F_{ij} \equiv (b^i, b^j)_z.
\eeq
The variances of these estimators are \cite{Jungman_1996}
\beq
\sigma_{\hat{f}_i}^2 = (F^{-1})_{ii},
\eeq
and the signal-to-noise ratio (SNR) for $f_{\rm NL}$ is $f_{\rm NL}/\sqrt{(F^{-1})_{00}}$.

\begin{figure}[htbp!]
\centering
\includegraphics[width=85mm]{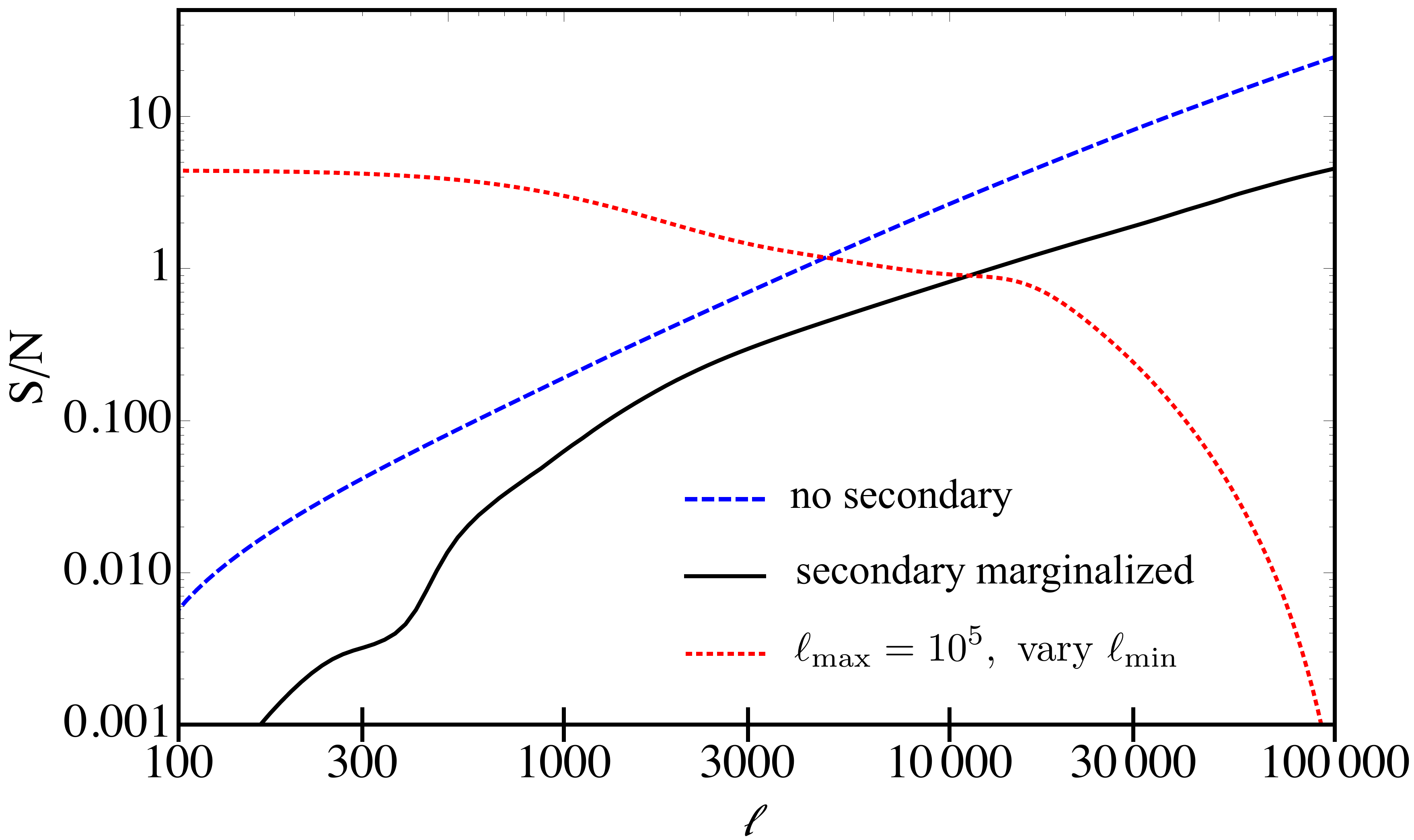}
\caption{Signal-to-noise ratio (SNR) for PNG of the local type with $f_{\rm NL} = 1$, for a single narrow redshift slice at $z = 50$ and assuming $f_{\rm sky} = 1$. The blue dashed curve shows $(F_{00})^{1/2}$, the SNR obtained if one neglected secondary non-gaussianity. The black solid and red dotted curves show $[(F^{-1})_{00}]^{-1/2}$, the SNR after marginalization over the unknown residual amplitudes of the secondary bispectrum, as a function of $\ell_{\max}$ (black solid) and as a function of $\ell_{\min}$ at fixed $\ell_{\max} = 10^5$ (red dotted).} 
\label{fig:snlo}
\end{figure}

We show in Fig.~\ref{fig:snlo} the forecasted SNR for the local-type PNG, for a single narrow redshift slice around $z = 50$, as a function of the maximum multipole moment $\ell_{\max}$ (with $\ell_{\min} = 100$). We also show for reference the SNR one would obtain if one neglected the secondary non-gaussianities, i.e.~when substituting $(F^{-1})_{00} \rightarrow 1/F_{00}$ as in Ref.~\cite{astro-ph/0611126}. We see that properly accounting for secondary non-gaussianities and their correlation with the primordial bispectrum reduces the SNR by a factor of $\sim 6$.

We also show the SNR integrated starting from $\ell_{\max}=10^5$ down to a minimum $\ell_{\min}$, as a function of the latter. 
It plateaus for $\ell_{\min}\sim10^3$, so modes with smaller $\ell$ do not contribute significantly to the signal-to-noise ratio, which justifies our neglect of several contributions to the bispectrum on large scales.

\begin{figure}[htbp!]
\centering
\includegraphics[width=85mm]{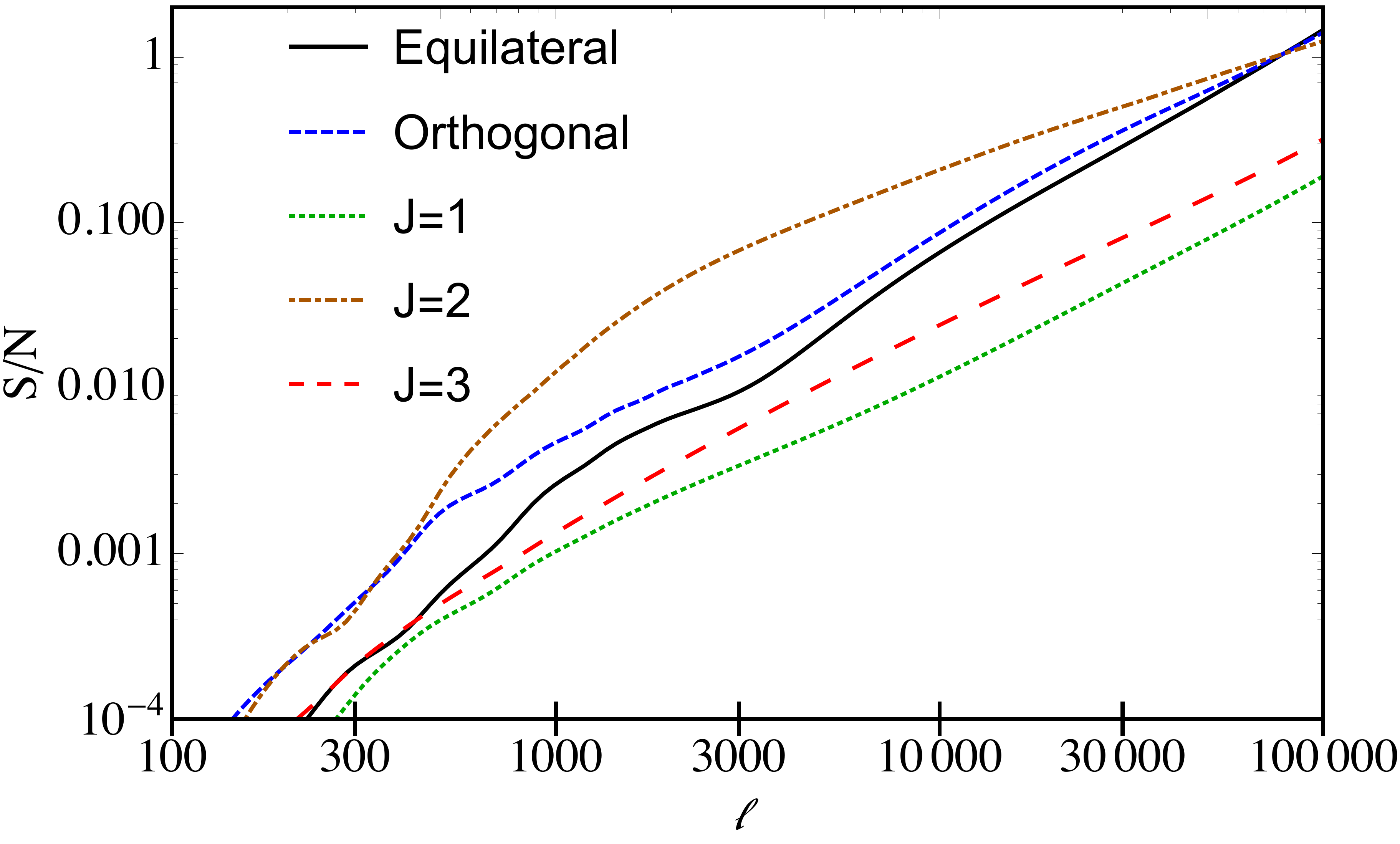}
\caption{SNR for different shapes of PNGs (with $\f = 1$), after marginalization over the residual amplitudes of the secondary bispectrum. The different lines correspond to equilateral-type PNG (solid black), orthogonal-type PNG (blue dashed), and the three direction-dependent shapes $J=1, 2$ and 3 in dotted green, dash-dotted brown, and long-dashed red, respectively.}
\label{fig:snrest}
\end{figure}

In Fig.~\ref{fig:snrest} we show the forecasted SNRs for the other shapes of PNG we considered. Secondary non-gaussianities are less correlated with these shapes than the they are with the local type, so the reduction in SNR is not as dramatic (a factor of $\sim 3$).

We summarize the forecasted SNR in Table \ref{tab:sn} for a single narrow redshift-slice at $z  = 50$, for $\ell_{\max} = 10^4$ (corresponding of an angular resolution of roughly 1 arcmin) and $\ell_{\max} = 10^5$ (0.1 arcmin angular resolution), assuming a cosmic-variance-limited experiment (i.e.~taking $C_{\ell}^{\rm tot} = C_{\ell}$, and neglecting additional thermal noise). In particular, we find that values of $f_{\rm NL}^{\rm loc} \sim 1.3$ and $\sim 0.23$ could be reached for $\ell_{\rm max} = 10^4$ and $10^5$, respectively. The bigger improvement for better resolution for the orthogonal and equilateral shapes with respect to the local one is due to the fact that they become less degenerate with the secondary bispectra as more modes are added in the analysis, as argued in Section~\ref{sec:bias}.


\begin{center}
\begin{table}[h]
    \begin{tabular}{| l | c | c | }
    \hline
    PNG type & $\sigma_{\f}$ (arcmin) & $\sigma_{\f}$ ($0.1$ arcmin) \\ \hline
    Local & 1.3 & 0.23  \\ 
    Equilateral  & 14 & 0.71 \\ 
    Orthogonal & 11 & 0.71 \\ 
    $J=1$ & 83 & 5.3  \\ 
    $J=2$ & 4.5 & 0.83  \\ 
    $J=3$ & 40 & 3.1  \\         
	\hline
    \end{tabular}
\caption{Detection forecasts for different shapes of PNG for a cosmic-variance-limited experiment observing the full sky at a single narrow redshift slice at $z=50$. The central column gives the results for $\ell_{\max}=10^4$ (equivalent to having an experiment with arcminute resolution), and the right column those for $\ell_{\max}=10^5$ (one tenth of arcminute).}
\label{tab:sn}
\end{table}
\end{center}

It is interesting to discuss how well we could probe the four secondary coefficients, $\overline{T}_{21}, \alpha, \beta$, and $\gamma$. In Tab.~\ref{tab:Cov} we show the relative errors reachable for each of them with a $0.1$ arcminute resolution, as well as the correlation with the rest of parameters.

\begin{center}
	\begin{table}[h]
		\begin{tabular}{| c | c | c | c | c | c | }
			\hline
			&  $\overline{T}_{21}$ &  $\alpha$ &  $\beta$ &  $\gamma$ & $f_{\rm NL}$ \\
			\hline
			$\overline{T}_{21}$& $5.9\times 10^{-4}$  &  &   &   &  \\
			$\alpha$ &  $-$0.95 & $2.6\times 10^{-3}$  &   &   &  \\
			$\beta$ &  $-$0.97 &  0.91 &  0.012 &   &  \\
			$\gamma$  &  0.41 &  $-$0.63 &  $-$0.43 &  $7.2\times 10^{-4}$ &  \\			   
			$f_{\rm NL}$ &  0.89 &  0.85 &  $-$0.92 & 0.36 & 0.23  \\
			\hline
		\end{tabular}
		\caption{Fractional error and correlation coefficients of $f_{\rm NL}$ and secondary bispectrum amplitudes. The diagonal elements are the fractional errors for each parameter, calculated as $\sqrt{(F^{-1})_{00}}$ for the case of $f_{\rm NL}$ and  $\sqrt{(F^{-1})_{ii}}/A_i^0$ for the rest.
				The off-diagonal elements are the correlations between parameters, defined as $(F^{-1})_{ij}/\sqrt{(F^{-1})_{ii}(F^{-1})_{jj}}$. For these results we considered local non gaussianity, at redshift $z=50$ and a resolution of 0.1 arcminutes.}
		\label{tab:Cov}
	\end{table}
\end{center}

\subsection{Choice of nuisance parameters} 
\label{App:B}

In our analysis we have marginalized over the residuals of the four coefficients $\overline{T}_{21}, \alpha, \beta, \gamma$. Here we discuss how different choices would affect our results.

On the optimistic side, if we were able to relate the four secondary coefficients to each other to high precision 
we could choose to marginalize over a single overall amplitude for the secondary bispectrum.

On the pessimistic side, we may choose to marginalize over all geometrically distinct contributions to the secondary bispectrum. This would account for unknown redshift dependences in the $c_i$ coefficients. Recalling that the kernels $F$ and $G$ are made of three geometrically distinct pieces, Eq.~(\ref{eq:secondary-grav}) gives 18 different geometric shapes. Equation \eqref{eq:secondary-nl} adds three independent shapes. This  amounts to a total of 21 distinct geometric shapes, the amplitudes of which we marginalize over.

\begin{figure}[h]
	\includegraphics[width=88mm]{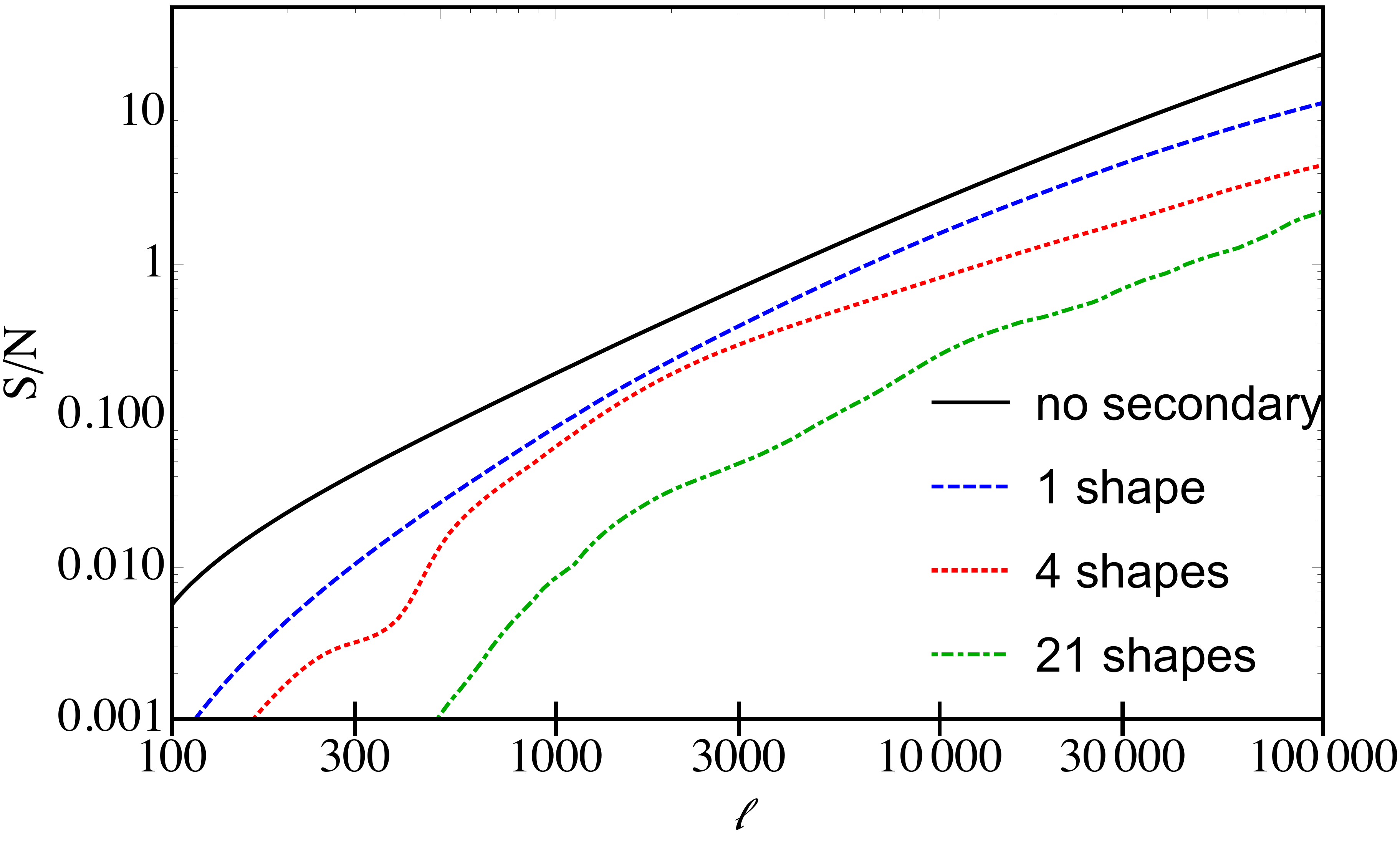}
	\caption{SNR for local-type PNG with $\f = 1$ as a function of $\ell_{\max}$, when neglecting secondary non-gaussianities (top, solid black curve), marginalizing over an overall amplitude of the secondary bispectrum (blue dashed), marginalizing over 4 coefficients as we do in the main text (red, dashed), and marginalizing over the amplitudes of the 21 geometrically distinct shapes of secondary bispectra (bottom, green dash-dotted).} 
	\label{fig:snbf}
\end{figure}

We show the resulting SNRs in Fig.~\ref{fig:snbf}, where for reference we also show the SNR in the absence of secondary non-gaussianities, and our main result, which considers 4 nuisance parameters. As expected, our result lies between the optimistic and pessimistic cases, which act as bounds for the SNR when considering additional secondary bispectra. 

In particular, in the optimistic approach, the SNR is improved by a factor of $\sim 5$: we find detection limits $\f^{\rm local}\sim 0.12$, $\f^{\rm equil}\sim 0.75$, $\f^{\rm ortho}\sim 0.58$, $\f^{J=1}\sim 5.7$, $\f^{J=2}\sim 0.88$, $\f^{J=3}\sim 13$ at arcminute resolution and $\f^{\rm local}\sim 0.0063$, $\f^{\rm equil}\sim 0.032$, $\f^{\rm ortho}\sim 0.030$, $\f^{J=1}\sim 0.19$, $\f^{ J=2}\sim 0.04$, $\f^{ J=3}\sim 0.40$ at maximum resolution for a single redshift slice at $z=50$. 

In a real experiment, a $\chi^2$-like test should be carried out to find out whether additional secondary bispectra to the four proposed here need to be considered.

\subsection{Tomography}

So far we have been studying the bispectrum on a single redshift slice, which would correspond to observing the 21-cm line with a single frequency channel. However, one of the great advantages of the 21-cm line is that it enables us to coadd information from different redshifts.

Before thinking of how to add different redshift shells we will study whether they contain the same or different information. Let us construct a measure of the correlation between two slices at a radial distance $\Delta r$ from each other. We define the correlation length $\xi_r(\ell)$ as the radial separation beyond which the cross-correlation between two redshift-slices is less than 1/2 the power spectrum:
\be
C_\ell[\Delta r = \xi(\ell)] = \dfrac 1 2 C_\ell[\Delta r = 0],
\ee
where the cross-power spectrum $C_{\ell}[\Delta r]$ is obtained from
\be
C_{\ell} [\Delta r]\equiv \frac1{r^2} \int \frac{d k_{||}}{2 \pi} P_{\delta T}\left(k_{||}, \bsl/r\right) e^{i k_{||} \Delta r}.
\ee

\begin{figure}[htbp!]
\centering
\includegraphics[width=85mm]{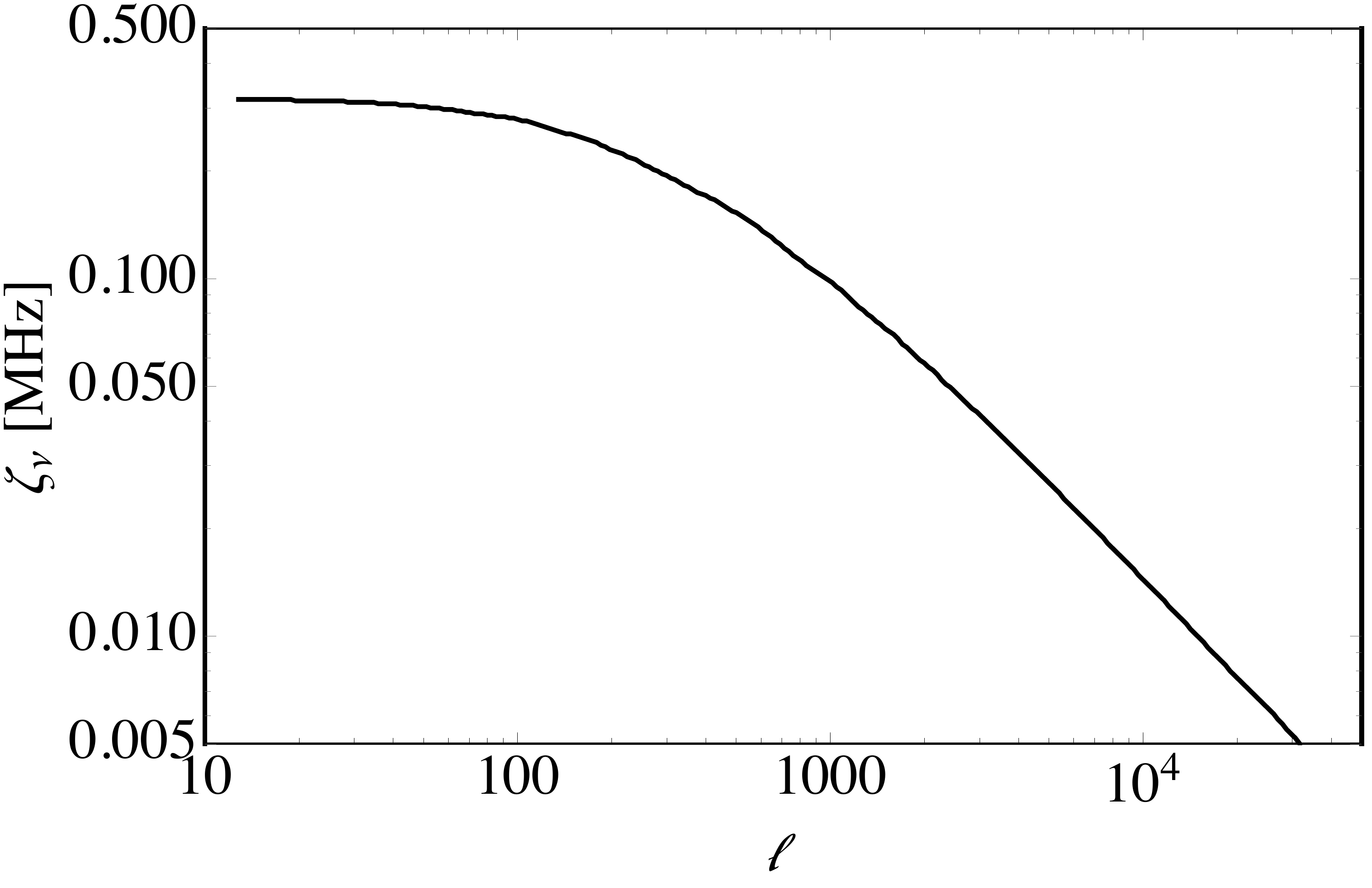}
\caption{Correlation length $\xi_\nu$ as a function of $\ell$, defined as the separation in frequency beyond which two redshift slices are correlated by less than 1/2. This curve was calculated with an infinitely narrow bandwidth at $z=50$, and for each $\ell$ the correlation would increase to match the value of the bandwidth if it is bigger than the $\xi_\nu$ in the plot.}
\label{fig:correlation}
\end{figure} 

From the correlation length in radial comoving separation $\xi_r$ we obtain the characteristic correlation length in frequency space $\xi_{\nu}$ through 
\barr
\xi_{\nu} &=& \dfrac{\mathrm d\nu}{\mathrm d z} \dfrac{\mathrm d z}{\mathrm dr} \xi_r = \nu_0 H_0 \sqrt{\Omega_M} (1+z)^{-1/2} \xi_r\\
&\approx& 1 \textrm{ MHz}  \left(\frac{51}{1+z}\right)^{1/2} \left(\dfrac{\xi_r}{60 ~\mathrm{ Mpc}} \right),  
\earr
where $\nu_0 = 1.4$ GHz is the rest-frame frequency of the 21-cm transition.  

We show the function $\xi_\nu(\ell)$ in Figure~\ref{fig:correlation}. For $\ell \lesssim 100$ (corresponding to $k \lesssim k_{\rm eq} \sim 0.01$ Mpc$^{-1}$), $P(k_{||}, \bs{\ell}/r)$ peaks at $k_{||} \sim k_{\rm eq}$, independently of $\ell$, and the cross-correlation $C_{\ell}[\Delta r]$ has a characteristic length scale $\xi_{\nu} \approx 0.3$ MHz, independent of $\ell$. For $\ell \gtrsim 100$, the function $P(k_{||}, \bs{\ell}/r)$ has a characteristic turnaround scale at $k_{||} \sim \ell/r$, which leads to a correlation length $\xi_{\nu}(\ell) \propto 1/\ell$.

In order to compare with previous results in the literature \cite{astro-ph/0610257, astro-ph/0611126} we will assume bandwidths $\Delta\nu$ of 1 MHz and 0.1 MHz. 
As argued above (Fig.~\ref{fig:snlo}) most of the signal comes from large-$\ell$ modes ($\ell\gtrsim 1000$), for which the correlation length $\xi_\nu < 0.1$ MHz, so in both cases we may assume that different redshift slices are completely uncorrelated. An observation of 21-cm fluctuations between 14 MHz ($z=100$) and 45 MHz ($z=30$) with frequency resolution $\Delta\nu$ would therefore have $N_{\rm slices} \approx 30 \times $ 1 MHz/$\Delta \nu$ independent redshift slices. 

The simplest analysis would consist in finding the best-fit $\f$ for each redshift slice and coadd the estimators with inverse-variance weighting. This procedure is not optimal, however, as the secondary bispectrum (and by extension, the residual after subtraction of the best-estimate $B^{\rm sec, 0}$) is a smooth function of redshift. The redshift dependence of the residuals $f_i = \Delta A_i$ can therefore be modeled by a linear combination of a few basis functions and depends on a few coefficients instead of $N_{\rm slices}$ independent amplitudes:
\beq
f_i(z) = \sum_{j = 0}^{N_{\rm bases}} f_{ij} P_j(z).
\eeq 
Several choices of basis functions could be made. We found that in the redshift range 30$-$100 the coefficients $A_i(z)$ could be fit to $\sim$ 1, 0.1, and 0.01 percent accuracy with third, fifth, or seventh-order polynomials in $\log(z)$, respectively. We assume that this will also hold for the residuals. We therefore adopt $P_j(z) = [\log(z/50)]^j$ for $j = 0$ to $N_{\rm bases}=3$ or 7 as our basis set. Our full model for the redshift-dependent bispectrum is therefore
\barr
B_{\ell_1 \ell_2 \ell_3}(z) &=& B_{\ell_1 \ell_2 \ell_3}^{\rm sec, 0}(z) + f_{\rm NL} b_{\ell_1 \ell_2 \ell_3}^{\rm prim}(z)\nonumber\\
 &+& \sum_{i = 1}^4 \sum_{j = 0}^{N_{\rm bases}-1} f_{ij} b^{(ij)}_{\ell_1 \ell_2 \ell_3}(z),
\earr
where $b^{(ij)}_{\ell_1 \ell_2 \ell_3}(z) \equiv P_j(z) \times b^i_{\ell_1 \ell_2 \ell_3}(z)$.

We now fit simultaneously for $f_{\rm NL}$ and $4\times N_{\rm bases}$ nuisance parameters $f_{ij}$. Because we assume the redshift slices are uncorrelated (specifically, the noise is uncorrelated in different slices, but the signal is not), the total scalar product between two bispectra is simply obtained by summing the single-redshift scalar product over redshift slices:
\beq
(b^n, b^m) \equiv \sum_z (b^n, b^m)_z, \label{eq:scal-tot}
\eeq
where $n \equiv (ij)$ is a generalized index, and $(b^n, b^m)_z$ is the scalar product of two bispectra at redshift $z$ defined in Eq.~\eqref{eq:scal-prod}. The usual Fisher analysis leads to $\sigma_{\hat{f}_{\rm NL}}^2 = (F^{-1})_{00}$, where the $(1 + 4 N_{\rm bases}) \times (1 + 4 N_{\rm bases}) $ Fisher matrix $F_{nm}$ is now defined from the total scalar product \eqref{eq:scal-tot}. We find that using seventh-order instead of third-order polynomials degrades the SNR by no more than $\sim 20$ \%. The final results we quote are obtained using third-order polynomials.
 
Our final results are shown in Table~\ref{tab:fnl}, where we quote the minimum $\f$ detectable for $\fsky=1$ and $\ell_{\max} = 10^5$ for two different bandwidths ($\Delta \nu =$1 and 0.1 MHz). For $f_{\rm sky} < 1$ all the results scale as $\sigma_{\f}\propto\fsky^{-1}$. 

In summary, with a bandwidth of 1 MHz we could cross the $\f=O(1)$ threshold, enabling us to rule out a big class of models of inflation if no PNG is detected. 
Increasing the frequency resolution to 0.1 MHz the numbers improve to $\f \sim \textrm{few}~ 10^{-2}$, which would be close to the ultimate limit of the consistency relation ($\f\sim n_s-1$), and hence should be present even in the simplest model of inflation.

\begin{center}
\begin{table}[h]
    \begin{tabular}{| l | c | c | }
    \hline
    PNG type & $\sigma_{\f}$ (1 MHz) & $\sigma_{\f}$ ($0.1$ MHz) \\ \hline
    Local & 0.12 & 0.03  \\ 
    Equilateral  & 0.39 & 0.04 \\ 
    Orthogonal & 0.29 & 0.03  \\ 
    $J=1$ & 1.1 & 0.1  \\ 
    $J=2$ & 0.33 & 0.05  \\ 
    $J=3$ & 0.85 & 0.09  \\         
	\hline
    \end{tabular}
\caption{Minimum $\f$ detectable integrating all redshift slices between $z=30$ and $z=100$ for $\fsky=1$. In the central column we show the result for a bandwidth of $\Delta\nu=1$ MHz and in the right column for  $\Delta\nu=0.1$ MHz.}
\label{tab:fnl}
\end{table}
\end{center}

\section{Conclusions}

Now that the information from the CMB on non-gaussianity has been almost fully mined, it is time to consider other potential data sets. Intensity fluctuations in the 21-cm line during the dark ages offer a window into yet unexplored times and scales, and a promising future probe of PNGs. 

The technical challenges that need to be overcome before the required experiments see the light of day are daunting. Because of atmospheric attenuation it would require an observatory on the Moon. Even then, care should be taken with intense Galactic foreground emission. Nevertheless, this is not an impossible task.

An additional issue is that the 21-cm signal is intrinsically highly non-gaussian, due to non-linear gravitational growth, and the non-linear mapping between brightness temperature and the underlying density field. In this paper we have, for the first time, addressed this issue with a rigorous Fisher analysis approach, assuming cosmic-variance limited experiments with a finite angular and frequency resolution. We have shown that for a single redshift slice the secondary bispectrum is significantly degenerate with the primordial one, which results in a noticeable decrease of the forecasted signal-to-noise ratio (SNR) for PNGs. This contrasts with the results of previous work, where this degeneracy was either neglected when forecasting the SNR \cite{astro-ph/0611126}, or where it was claimed to be weak \cite{astro-ph/0610257}. We then co-added the information of independent redshift slices while enforcing a smooth variation of the secondary bispectrum amplitudes with redshift.

For a full-sky experiment with $\Delta \nu = 0.1$ MHz and 0.1-arcminute resolution, we forecast a sensitivity $\sigma_{\f^{\rm local}}\approx 0.03$, which would enable us to check the famous inflationary consistency relation. We also forecast $\sigma_{f^{\rm equil}}\approx 0.04$, $\sigma_{f_{\rm ortho}} \approx 0.04$, $f^{\rm J = 1} \approx 0.1$, $f^{\rm J = 2} \approx 0.05$, and $f^{\rm J = 3} \approx 0.09$. Measurements of 21-cm fluctuations therefore have the potential to significantly improve upon cosmic-variance-limited CMB bounds.

\begin{acknowledgments}
We thank Ely Kovetz, Nikhil Anand, and Alvise Raccanelli for useful discussions.
This work was supported by NSF Grant No. 0244990, NASA NNX15AB18G, the John Templeton Foundation, and the Simons Foundation.
\end{acknowledgments}

\end{document}